\documentclass[pra,aps,reprint,a4paper,superscriptaddress,longbibliography,floatfix]{revtex4-2}
\usepackage{graphicx}
\usepackage{epsfig}
\usepackage{amsfonts}
\usepackage{amsmath}
\usepackage{amssymb}
\usepackage{color}
\usepackage{comment}
\usepackage[colorlinks=true,linkcolor=blue,citecolor=magenta,urlcolor=blue]{hyperref}

\newtheorem{prop}{Proposition}
\newtheorem{lem}[prop]{Lemma}
\newtheorem{theo}[prop]{Theorem}
\newtheorem{cor}[prop]{Corollary}

\def\proof#1{\noindent{\em Proof.} #1\hfill $\Box$\medskip}

\begin{document}


\title{Irreducible Magic Sets for $n$-Qubit Systems}


\author{Stefan Trandafir}
\email{stefan_trandafir@sfu.ca}
\affiliation{Department of Mathematics, Simon Fraser University, Burnaby, British Columbia, V5A 1S6, Canada}

\author{Petr Lison\v{e}k}
\email{plisonek@sfu.ca}
\affiliation{Department of Mathematics, Simon Fraser University, Burnaby, BC, V5A 1S6, Canada}

\author{Ad\'an~Cabello}
\email{adan@us.es}
\affiliation{Departamento de F\'{\i}sica Aplicada II,
Universidad de Sevilla, E-41012 Sevilla, Spain}
\affiliation{Instituto Carlos~I de F\'{\i}sica Te\'orica y Computacional, Universidad de
Sevilla, E-41012 Sevilla, Spain}


\begin{abstract}
Magic sets of observables are minimal structures that capture quantum state-independent advantage for systems of $n\ge 2$ qubits and are, therefore, fundamental tools for investigating the interface between classical and quantum physics. A theorem by Arkhipov (arXiv:1209.3819) states that $n$-qubit magic sets in which each observable is in exactly two subsets of compatible observables can be reduced either to the two-qubit magic square or the three-qubit magic pentagram [N.\ D.\ Mermin, Phys.\ Rev.\ Lett.\ \textbf{65}, 3373 (1990)]. An open question is whether there are magic sets that cannot be reduced to the square or the pentagram. If they exist, a second key question is whether they require $n >3$ qubits, since, if this is the case, these magic sets would capture minimal state-independent quantum advantage that is specific for $n$-qubit systems with specific values of $n$. Here, we answer both questions affirmatively. We identify magic sets that cannot be reduced to the square or the pentagram and require $n=3,4,5$, or $6$ qubits. In addition, we prove a generalized version of Arkhipov's theorem providing an efficient algorithm for, given a hypergraph, deciding whether or not it can accommodate a magic set, and solve another open problem, namely, given a magic set, obtaining the tight bound of its associated noncontextuality inequality.
\end{abstract}


\maketitle


{\em Introduction.---}A magic set for a system of $n\ge 2$ qubits \cite{Peres90,Peres91,Peres92,Mermin90,Mermin93} is a set of Pauli observables (i.e., those represented by $n$-fold tensor products of single-qubit Pauli operators $I$, $X$, $Y$, and $Z$) and contexts (subsets of compatible observables represented by commuting operators and such that their product is the identity---in the case of ``positive'' contexts---or minus the identity---in the case of ``negative'' contexts---) with the following properties:
(i)~each observable is in an even number of contexts.
(ii)~The number of negative contexts is odd. 
(iii)~The set is minimal: properties (i) and (ii) do not hold if any observable is removed. As a simple parity argument shows, properties (i) and (ii) make it impossible to assign a predetermined outcome, either $1$ or $-1$, to each observable while satisfying that the product of the outcomes for the observables of a positive (negative) context is $1$ ($-1$), as predicted by quantum mechanics (QM). Consequently, any magic set provides a simple state-independent proof of the impossibility of simulating QM with noncontextual hidden variable (NCHV) models \cite{Peres90,Peres91,Peres92,Mermin90,Mermin93,Cabello08}.

In addition, the most famous magic sets have a fourth property:
(iv)~their hypergraph of compatibility (i.e., the one in which each vertex represents an observable and each hyperedge a context) is vertex-transitive (i.e., its automorphism group acts transitively on its vertices). A hypergraph $H = (V, E)$ is a finite set $V$ of vertices and a finite set $E$ of hyperedges, where each hyperedge is a multiset of vertices. Besides symmetry and elegance, vertex transitivity is helpful for experimental purposes.

\begin{figure}[t!]
\includegraphics[width=7.9cm]{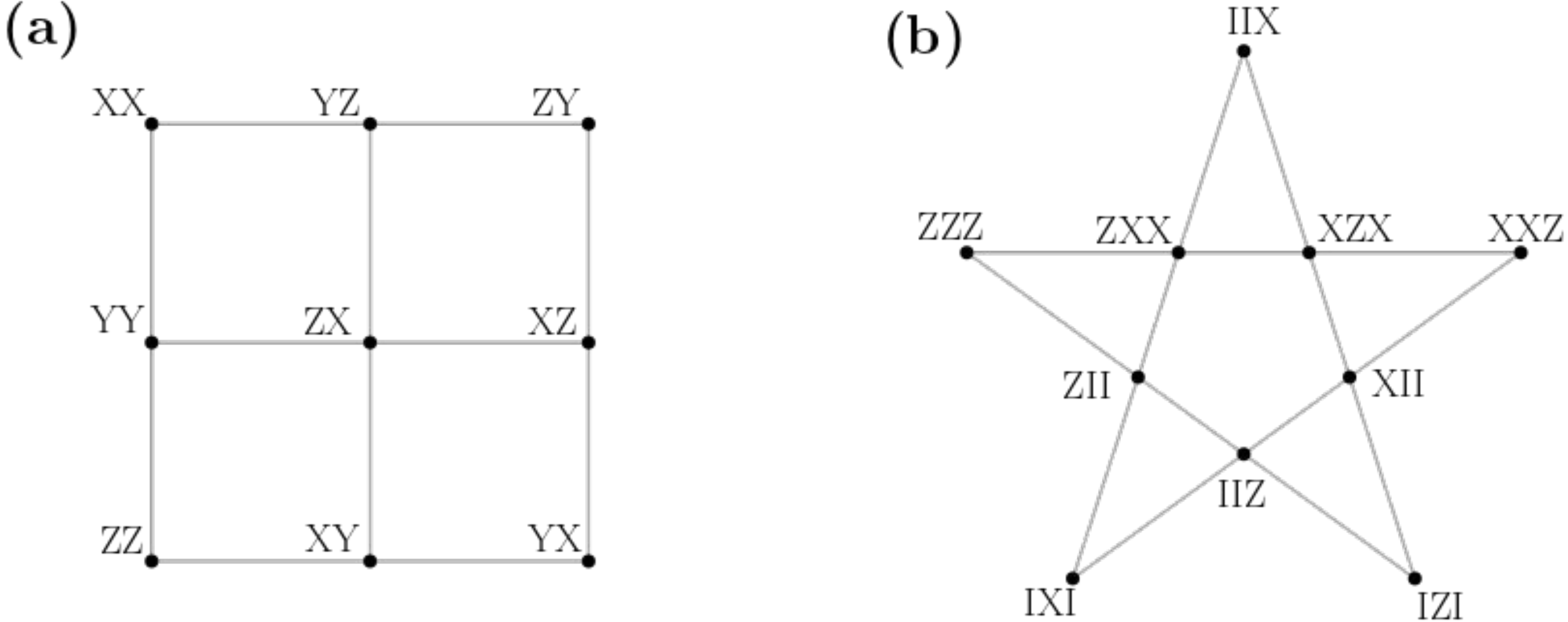}
\caption{(a) The magic square. (b) The magic pentagram. Each dot represents a Pauli observable.
$XZ$ denotes the observable represented by $\sigma_x \otimes \sigma_z$. $IXI$ denotes the observable represented by $\openone \otimes \sigma_x \otimes \openone$, where $\openone$ is the $2 \times 2$ identity matrix. Observables in the same straight line are mutually compatible and the product of their operators is the identity, except for the three vertical lines in~(a) and the horizontal line in~(b), where it is minus the identity.} 
\label{Fig1}
\end{figure}

There are two famous magic sets. One is the ``magic square,'' ``Peres-Mermin table,'' or ``Mermin square'' for $n=2$ qubits found \cite{history} by Peres \cite{Peres90,Peres91,Peres92} and Mermin \cite{Mermin90,Mermin93} and shown in Fig.~\ref{Fig1}(a). The other is the ``magic pentagram'' or ``Mermin's star'' for $n=3$ qubits found by Mermin \cite{Mermin90,Mermin93} and shown in Fig.~\ref{Fig1}(b). Both sets were introduced as simplified proofs of the Kochen-Specker theorem \cite{KS67}. The adjective ``magic'' was first used in \cite{Aravind04}.

Magic sets have multiple applications (for details, see \cite{SM}), including Greenberger-Horne-Zeilinger-like proofs with two observers \cite{Cabello01b}, bipartite Bell inequalities with maximal quantum violation saturating the nonsignaling bound \cite{Cabello01b,CBPMD05,YZZYZZCP05,GMS07,AGA12}, obtaining Kochen-Specker sets of rays \cite{Peres91,DP97}, nonlocal games \cite{Aravind02,Aravind04,BBT05}, state-independent noncontextuality inequalities \cite{Cabello08,KZG09,ARBC09,MRCL10,Cabello10b}, measurement-based quantum computation \cite{AB09,Raussendorf13,DGBR15,RBDOB17}, nonlocality based on local contextuality \cite{Cabello10,LHC16,Cabello21}, device-independent quantum key distribution \cite{HHHHPB10,JMS20}, memory cost of classically simulating sequences of quantum measurements \cite{KGPLC11,FK17,CGGX18}, state-independent quantum dimension witnessing \cite{GBCKL14}, entropic inequalities \cite{RKK15}, device-independent self-testing \cite{WBMS16,KM17,CS17,CMMN20}, and quantum gravity \cite{LHS17}.


In a nutshell, the importance of magic sets lies in the fact that they are minimal structures that capture quantum state-independent advantage for an $n$-qubit system and thus are fundamental tools for investigating the interface between classical and quantum physics \cite{SM}.

Magic sets are useful to capture the quantum advantage. But the quantum advantage grows with $n$. Therefore, an interesting question is whether there are magic sets for $n > 3$ and how they are related to those for smaller values of $n$. A theorem by Arkhipov \cite{Arkhipov12} 
(see also \cite{LRS14}) suggests that the cases $n=2$ and $n=3$ are special. Arkhipov's theorem states that the intersection graph of the contexts of any magic set in which each observable is in exactly two contexts must contain either the intersection graph of the contexts of the magic square or the magic pentagram. The intersection graph of a family of sets is a graph in which each set is represented by a vertex and edges connect intersecting sets. 
A consequence of Arkhipov's theorem is that ``the magic square and magic pentagram are `universal' for magic games'' in which each observable is in exactly two contexts \cite{Arkhipov12}. A second consequence is that the magic sets with $n > 3$ qubits described in the literature \cite{PRC91,SP12,Planat12,Planat13,WA13,WA13b,Waegell14} derive from the square and the pentagram. However, Arkhipov's theorem leaves open some key questions:

(1) For $n=2$ qubits, each Pauli observable can be only in three contexts. Therefore, for $n=2$, the only even number that can be used to define magic sets following condition (i) is two. But this is not true for $n \ge 3$ qubits. Does the conclusion of Arkhipov's theorem hold if the requirement of each observable being in exactly two contexts is replaced by the requirement of each observable being in an even number $m$ of contexts? Are there, in this more general case, magic sets that cannot be reduced to the square and the pentagram?

(2) If the answer to the second question in (1) is affirmative, are there magic sets that cannot be reduced to any magic set with $n=2$ or $n=3$ qubits and thus are genuine to systems of $n > 3$ qubits? This is important as it would identify fundamental structures that are genuine for a specific number of qubits and thus can be used to certify whether a system has at least $n$ qubits.

(3) If the answer to (2) is affirmative, how does one identify those magic sets? Is it possible to generalize Arkhipov's theorem (which is essentially an efficient algorithm to check whether or not a hypergraph can accommodate a magic set under the assumption that each observable is in exactly two contexts) while removing the extra assumption?

All these questions seem to be important and, collectively, can be rephrased as follows: are there simple tools to detect and quantify quantum computational advantage for $n$-qubit systems that are specific for each value of $n$ and have gone unnoticed?
In this Letter, we answer all these questions in the affirmative.

Any magic set provides a logical contradiction between QM and NCHV models. However, translating that contradiction to an experiment requires deriving a noncontextuality inequality \cite{Cabello08} that is violated (for any initial state) measuring the elements of the magic set. There exists a general method for, given a magic set, obtaining a contextuality witness \cite{Cabello10}. Calculating the quantum value of that witness is immediate. Calculating its maximum for NCHV models is straightforward if the magic set is small. However, an open problem is obtaining the bound for NCHV models in general. In this Letter, we also solve this problem.


\begin{figure*}[hbt!]
\includegraphics[width=5.5cm]{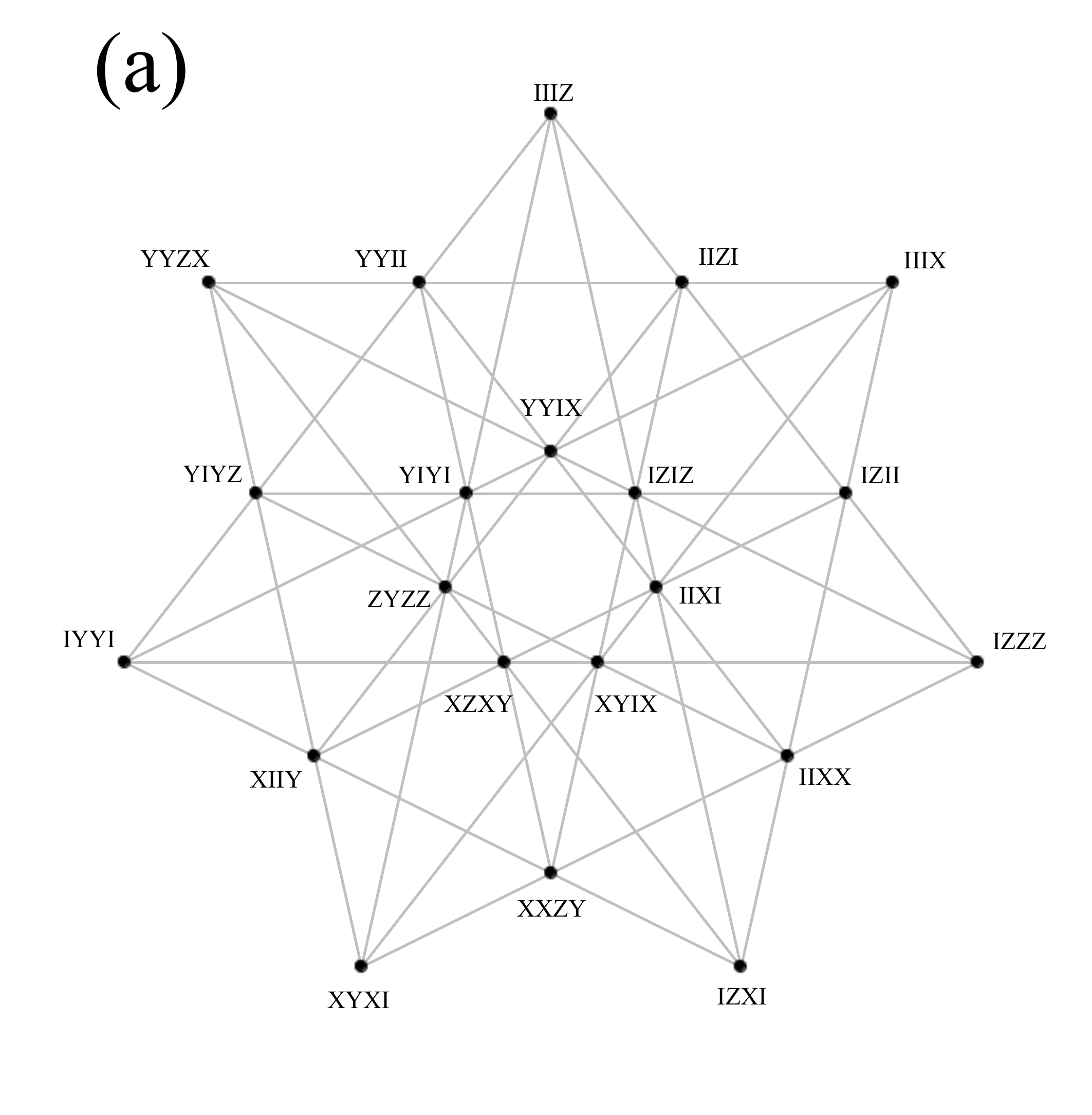}
\includegraphics[width=6.1cm]{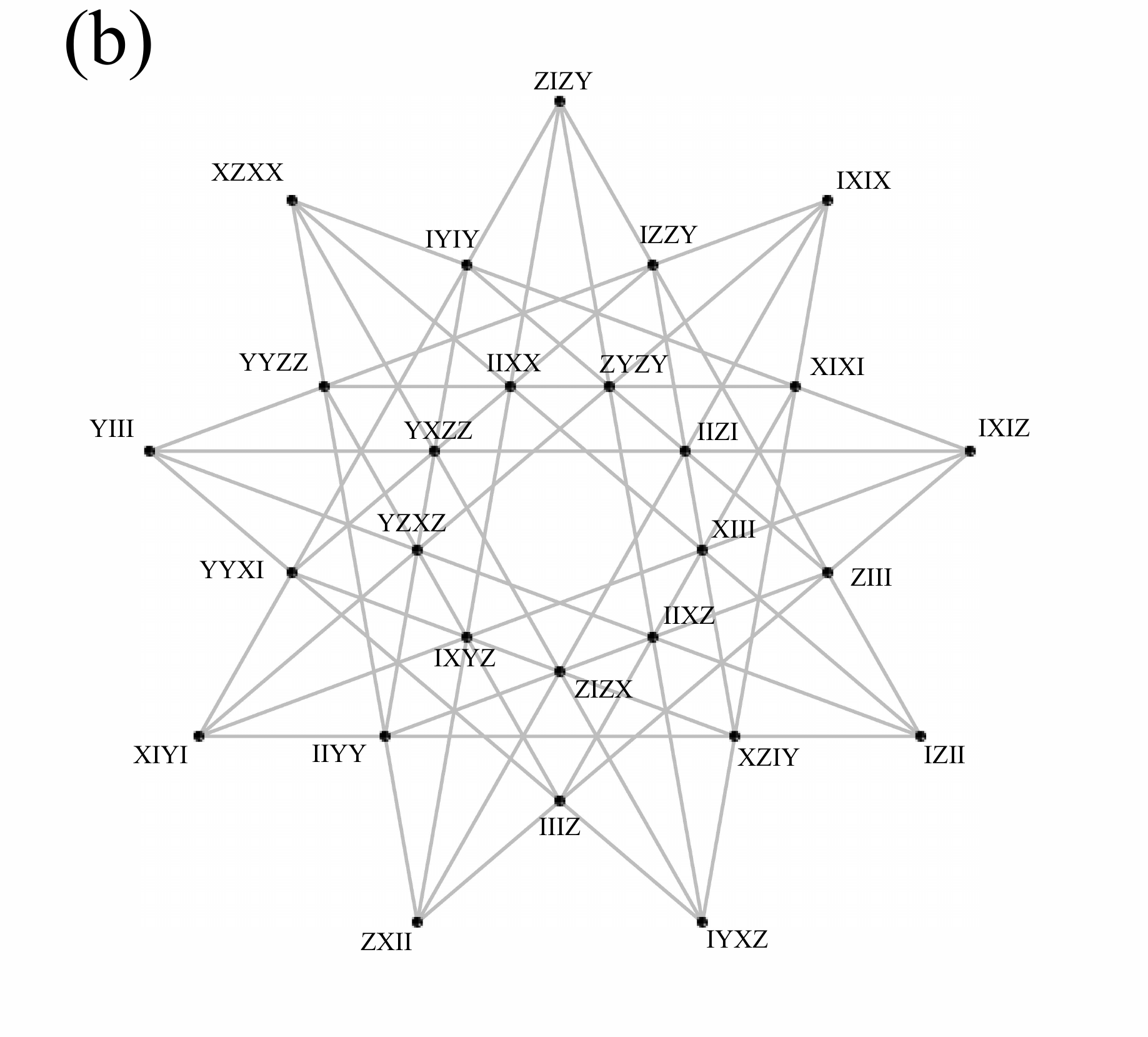}
\includegraphics[width=6.1cm]{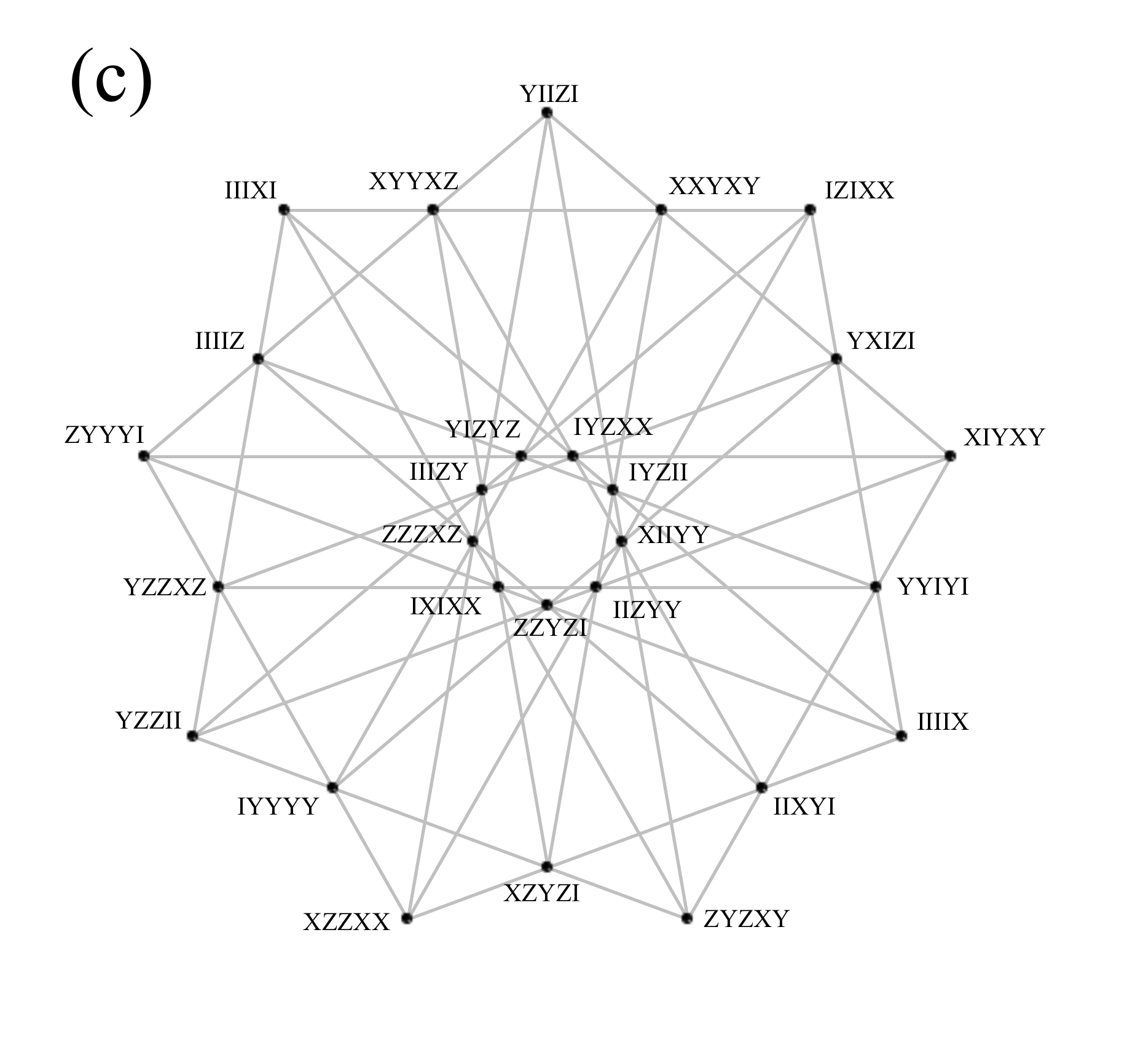}
\caption{The three magic sets with vertex-transitive graphs of compatibility with straight line representations in the Euclidean plane that cannot be reduced to the square or pentagram that we have found in this work. The notation is the same as that used in Fig.~\ref{Fig1}. (a)~MS4-21 requires $n=4$ qubits and has $21$ observables and $21$ contexts. In the example shown, $3$ of the contexts are negative. 
(b)~MS4-27 requires $n=4$ qubits and has $27$ observables and $27$ contexts. In the example shown, $5$ of them are negative.
(c)~MS5-27 requires $n=5$ qubits and has $27$ observables and $27$ contexts. In the example shown, $13$ of them are negative.
} 
\label{Fig2}
\end{figure*}


{\em Methodology.---}Finding all magic sets for any $n>3$ is intractable. There are $4^n-1$ Pauli observables, each of them is in~$\prod_{j=1}^{n-1} (1+2^j)$ positive or negative contexts of $2^n-1$~elements. These contexts of maximal size contain subsets whose product is the identity or minus the identity (e.g., $\{IIX,IXI,XII,XXX,IXX,XIX,XXI\}$ contains two subsets whose product is the identity: $\{IIX,IXI,XII,XXX\}$ and $\{IXX,XIX,XXI\}$). Unlike the case $n=2$, where $m$ can only be $2$ (since, each observable is in exactly $3$~contexts of maximal size), the possible values for $m$ grow with $n$.

However, since our main motivation is answering whether or not there are magic sets not covered by Arkhipov's theorem, we restrict our computational search to magic sets in which each observable is in four~contexts (the simplest case not covered by Arkhipov's theorem) and assume that contexts have four or five observables. The theoretical results presented in this Letter do not require these assumptions.

In addition, we use the following observation. Given a magic set $S$, each Pauli observable $o \in S$ can be represented by a vertex $v\in V$ and each context by an hyperedge $e \in E$ of its hypergraph of compatibility $H = (V, E)$. For example, Figs.~\ref{Fig1}(a) and (b) show $H$ for the magic square and pentagram, respectively (representing vertices by dots and hyperedges by straight lines connecting several dots). For a fixed $n$, there are different sets of Pauli observables whose relations of compatibility are represented by the same $H$. We say that two magic sets belong to the same class if they have the same $H$. 
For example, for $n=2$ qubits, there are $10$~magic sets sharing the hypergraph $H$ shown in Fig.~\ref{Fig1}(a).
Our strategy for finding magic sets is thus based on identifying hypergraphs $H$ that can represent magic sets. Specifically, we use the following algorithm.
(a) We fix the number of observables, say $N$,
in the putative magic set. 
We then use the list of groups acting transitively on $N$ points provided by computer algebra systems such as GAP \cite{GAP} or \textsc{Magma} \cite{Magma}. For each group $G$, we generate the orbits of $G$ acting on the subsets of $\{1,\ldots,N\}$ of size~$s$, where
$s\in\{4,5\}$. If any such orbit contains exactly $4N/s$ sets, then, by a simple counting argument, these sets are the hyperedges of a vertex-transitive 
hypergraph $H$ in which each vertex is in four edges. 
(b) We then use a theorem (Theorem~7 in \cite{SM}) to determine whether $H$ admits a magic assignment
of its vertices by $n$-qubit Pauli operators
for some $n$ (i.e., an $n$-qubit magic set).
If it does, then we also determine the smallest
such~$n$. We can also iterate through all such assignments.
(c) Whenever we find structures that are not minimal [i.e., which do not satisfy (iii)], we can find new structures that are minimal by a method detailed in~\cite{SM}, which also contains further details on the whole algorithm. We can also compute the minimum number of qubits needed and assignments in this case. The examples we find from this procedure need not be vertex-transitive and may have contexts of larger size and observables in a larger number of contexts.


{\em Results.---}With the assumptions made above, it can be seen that $H$ must have $N \ge 13$ vertices. By exhaustive computer search, we have found that there are no magic sets with fewer than $19$ vertices (Pauli observables), even if we drop the requirement that the hypergraph is vertex-transitive. We have also found that there are no magic sets with fewer than $20$ vertices that have at least one nontrivial automorphism.

\begin{figure}[htb!]
\includegraphics[width=6.2cm]{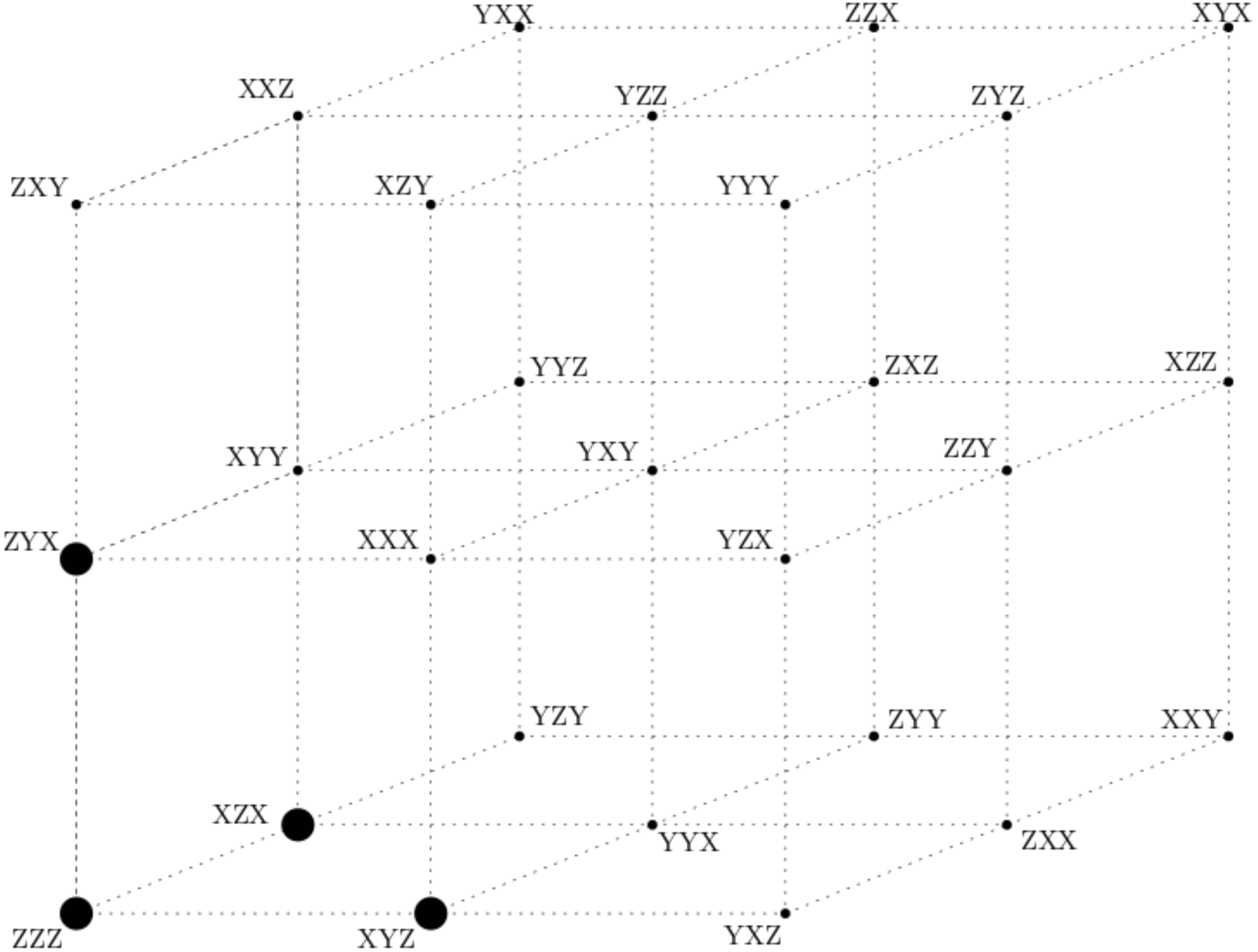}
\caption{The fourth magic set with vertex-transitive graph of compatibility that cannot be reduced to the square or pentagram that we have found, MS3-27, requires $n=3$ qubits and has $27$~observables and $27$~contexts. It does not admit a straight line representation in the Euclidean plane. Here, observables are given by coordinate points $(x,y,z) \in \mathbb{Z}_3^3$ (the $x$ axis is horizontal, the $y$ axis goes into the page, and the $z$ axis is vertical) and all contexts can be obtained by applying translations to a starter context. A possible starter context $\beta$ given by the observables corresponding to the coordinates $(0, 0, 0), (1, 0, 0), (0, 1, 0), (0, 0, 1)$ is depicted by the four larger dots.
All $27$~contexts can be generated by applying the translations $T_{a,b,c} : \mathbb{Z}_3^3 \to \mathbb{Z}_3^3,\ 0 \leq a,b,c \leq 2$ given by $(x,y,z) \to (x+a,y+b,z+c)$ to each of the observables of the starter context. For example, one obtains the context $\{ZYY, XXY, ZXZ, XYZ\}$ by applying the translation $T_{1,2,0}$ to the observables of $\beta$. In the example shown, all the contexts are negative.
The dotted edges appear only as a visual aid to make clear the correspondence of the vertices to the coordinates. Like the magic square, see Fig.~\ref{Fig1}(a), MS3-27 can be implemented using all $3^n$ $n$-qubit Pauli observables not containing $I$.}
\label{Fig3}
\end{figure}

We have found four classes of irreducible magic sets that have a vertex-transitive hypergraph of compatibility [i.e., that also satisfy property~(iv)] like the square and pentagram. Their hypergraphs and a magic assignment for each of them are presented in Figs.~\ref{Fig2}(a), (b), (c), and~\ref{Fig3}.

The one with the smallest number of observables is the class shown in Fig.~\ref{Fig2}(a), which requires $n=4$ qubits and has $21$~observables and $21$~contexts. Its $H$ is the so-called Gr\"unbaum-Rigby configuration \cite{GR90} already described by Klein \cite{Klein79}. 

Each of the other three classes has $27$~observables and $27$ contexts. The class in Fig.~\ref{Fig3} requires $n=3$ qubits.
The class in Fig.~\ref{Fig2}(b) requires $n=4$ qubits. Its $H$ is the 3-astral 4-configuration in \cite{Grunbaum09} [Fig.~3.7.2(b)]. The class in Fig.~\ref{Fig2}(c) requires $n=5$ qubits. Its $H$ is the smallest known weakly flag-transitive configuration \cite{Marusic99}.

The automorphism groups of the classes in Figs.~\ref{Fig2}(a)--(c) allow for straight line representations in the Euclidean plane (the ones shown in Fig.~\ref{Fig2}). However, such a representation is not possible for the class in Fig.~\ref{Fig3}. Instead, we can visualize its hypergraph by describing its automorphism group, as shown in Fig.~\ref{Fig3}.

We have also found irreducible magic sets not satisfying property (iv) (vertex-transitivity). They include one with $n=6$ qubits. See~\cite{SM} for details.


These sets by themselves answer question (1): there are magic sets that cannot be reduced to the square and the pentagram, including some that also satisfy property~(iv). They also answer question (2): there are magic sets that are genuine (irreducible to any magic set with a smaller number of qubits) to systems of $n=4$ [Figs.~\ref{Fig2}(a) and (b)], $n=5$ qubits [Fig.~\ref{Fig2}(c)], and $n=6$ qubits \cite{SM}.


{\em Extending Arkhipov's theorem.---}Here, we address question (3). Arkhipov's theorem provides an efficient algorithm to check whether or not an hypergraph yields a Pauli-based magic assignment satisfying that each observable is in exactly two contexts and the number of negative contexts is odd. The question is whether there is an efficient algorithm to check whether or not a hypergraph admits a Pauli-based magic assignment [i.e., can accommodate Pauli observables satisfying properties (i), (ii), and (iii)]. 

Steps (b) to (c) of our algorithm provide an efficient algorithm to check whether or not a hypergraph admits a Pauli-based magic assignment satisfying (i) and (ii). Therefore, in a sense, they answer question~(3). Additionally, step~(c) allows us to generate and iterate through magic assignments of minimal structures. The main result we exploit is the following theorem.

\begin{theo} 
\label{thm-magic-test}
Let $H$ be a proper Eulerian hypergraph with valid Gram space $\mathcal{V}$. Let $\mathcal{B}$ be any basis for $\mathcal{V}$. Then,
\begin{enumerate}
\item $H$ has a magic assignment with Pauli observables if and only if there is a magic Gram matrix in $\mathcal{B}$.
\item $H$ has a magic assignment with Pauli observables for a system of $k$ qubits satisfying (i) and (ii) if and only if there is a magic Gram matrix of binary rank at most $2k$ in $\mathcal{V}$.
\end{enumerate}
\end{theo}

A proper Eulerian hypergraph is a hypergraph with each vertex in an even number of distinct hyperedges. The valid Gram space is the set of $|V| \times |V|$ matrices $M$ whose entries satisfy the following linear equations: (a) $M_{i,j} = 0$ whenever vertices $v_i,v_j$ occur in the same hyperedge; and (b) $\sum\limits_{v_i \in e} M_{i,j} = 0$, for all $1 \leq j \leq |E|$ for all hyperedges $e \in E$. 
A magic assignment $\alpha : V \to GL(\mathcal{H})$, where $\mathcal{H}$ is a Hilbert space, is an assignment such that: (A) $ \alpha(v)^2 = I$ and $\alpha(v)$ is Hermitian for all $v\in V$.
(B) $\alpha(v)\alpha(w) = \alpha(w)\alpha(v)$ whenever $v, w$ are in a common hyperedge $e \in E$.
(C) $\prod\limits_{v\in e} \alpha(v) = \pm I$ for each hyperedge $e \in E$.
(D) $\prod\limits_{v \in e} \alpha(v) = -I$ for an odd number of hyperedges $e \in E$.

Using Arkhipov's result, our methodology yields a novel algorithm for checking graph planarity (\cite{SM}, Corollary~12). Additionally, in the case that the graph $G$ is nonplanar, this algorithm also produces a magic Gram matrix encoding a copy of $K_{3,3}$ or $K_5$ appearing as a topological minor of $G$. 


{\em Noncontextuality inequalities.---}Given a set of Pauli observables satisfying (i) and (ii) (i.e., not necessarily minimal), let us call $C_p$ its set of positive contexts and $C_n$ its set of negative contexts. Then, as shown for the square and the pentagram in \cite{Cabello08}, and for more general cases in \cite{Cabello10}, the following inequality must be satisfied by any NCHV model:
\begin{equation}
\label{nci}
\sum_{{\cal C}_i \in C_p} \langle {\cal C}_i \rangle - \sum_{{\cal C}_j \in C_n} \langle {\cal C}_j \rangle \le b,
\end{equation}
where $\langle {\cal C}_i \rangle$ denotes the mean value of the products of all the observables in context ${\cal C}_i$. QM makes a prediction for each context (that the product is either $-1$ or $1$). The limit for NCHV models is $b = 2s -|C_p| -|C_n|$, where $s$ is the maximum number of quantum predictions that can be simultaneously satisfied by a NCHV model \cite{Cabello10}. 
An open problem \cite{Cabello10} is, given a hypergraph $H$ corresponding to a magic set, what is $b$? Here we solve this problem in two senses. On the one hand, we give a method for computing $b$ by using results from coding theory (see~\cite{SM} for details). Computing $b$ is important for, e.g., computing the resistance to noise of the quantum advantage of any magic set \cite{Cabello10}. On the other hand, we prove a more general result.

\begin{theo} \label{theo:lowest-hamming-weight}
Let $H = (V, E)$ be a magic Eulerian hypergraph with incidence matrix $M$. Let $\alpha$ be a magic assignment of $H$, and let $w_{\min}$ be the minimum of
Hamming weights 
of elements of the affine space $c(\alpha) + \operatorname{row}(M)$. Then,
the noncontextual bound
for $\alpha$ is $b = |E| - 2w_{\min}$.
\end{theo}

Given a hypergraph $H = (V, E)$ with vertices $v_1, \dots, v_m$ and edges $e_1, \dots, e_n$, the incidence matrix of $H$ is the $m \times n$ binary matrix $M$ for which $M_{i,j} = 1$ whenever $v_i \in e_j$. By $\operatorname{row}(M)$ we denote the row space of the matrix $M$. The Hamming weight of a binary vector $w$ is the number of nonzero coordinates of $w$. Given a magic assignment $\alpha$ of $H$, we define $c(\alpha) \in GF(2)^{n}$ to be the vector for which $c(\alpha)_i = 0$ whenever $\prod_{v \in e_i} \alpha(v) = 1$ and $c(\alpha)_i = 1$ otherwise.


{\em Conclusions.---}Minimal vertex-transitive magic sets are fascinating objects used in a wide variety of areas as they capture minimal quantum state-independent advantage for $n$-qubit systems and are thus fundamental tools for investigating the interface between classical and quantum physics. While Arkhipov's theorem might have been taken as an indication that there are only two classes of irreducible vertex-transitive magic sets, one requiring two and the other requiring three qubits, and that all magic sets derive from them, in this Letter, we have shown that the landscape of magic sets is quite different from the one suggested by Arkhipov's theorem as there are, at least, four more classes: one requiring three qubits, here called MS3-27, that cannot be drawn in a plane (see Fig.~\ref{Fig3}); two requiring four qubits, here called MS4-21 [see Fig.~\ref{Fig2}(a)] and MS4-27 [see Fig.~\ref{Fig2}(b)]; and one requiring five qubits, MS5-27 [see Fig.~\ref{Fig2}(c)]. We have also found other irreducible magic sets requiring from three to six qubits (but not vertex-transitive ones). 

In the light of these results, it seems that each $n$ has its own set of irreducible vertex-transitive magic sets. Finding them and especially finding the ones with minimum number of observables (so far, the magic square for $n=2$, the magic pentagram for $n=3$, MS4-21 for $n=4$, and MS5-27 for $n=5$) is an interesting challenge for the reasons that have motivated this work (namely, identifying minimal structures providing state-independent quantum advantage and requiring a specific number of qubits). One possible way to obtain these sets would be by generalizing to a higher number of qubits the geometrical structure of the sequence pentagram, MS4-21, and MS5-27, as well as the sequence square and MS3-27.

In addition, we have proven a general expression for the classical (noncontextual) bound of the inequality associated to any magic state (minimal or not), which is useful for many purposes as it allows us, e.g., to compute the robustness to noise in the implementation of the Pauli observables (or, in general, versus any type of experimental limitation) for any given magic set \cite{SM}. We hope these results stimulate further research on magic sets and their applications.


\begin{acknowledgments}
We thank T.\ Pisanski for pointing out that the hypergraphs in Figs.~\ref{Fig2}(b) and (c) appeared in Refs.~\cite{Grunbaum09,Marusic99}, respectively. This work was supported by 
the Natural Sciences and Engineering Research Council of Canada (NSERC, Project No.\ RGPIN-2015-06250 and RGPIN-2022-04526),
Project Qdisc (Project No.\ US-15097, Universidad e Sevilla), with FEDER funds, QuantERA grant SECRET, by MINECO (Project No.\ PCI2019-111885-2), and MICINN (Project No.\ PID2020-113738GB-I00).
\end{acknowledgments}




\appendix

\section{Some applications of magic sets}


Why are magic sets important in physics? The purpose of this appendix is to guide the reader to some of the applications, uses, and connections in which magic sets are involved. The order is chronological and the list is not exhaustive. We just want to give an idea about the variety of problems in which magic sets play an important role.


\subsection{Magic sets and two-observer GHZ-like proofs}


In 1989, Greenberger, Horne, and Zeilinger (GHZ), provided a logical argument of impossibility of local hidden variables involving four spatially separated observers \cite{GHZ89}. Later on, Mermin \cite{Mermin90} provided a three-observer version of it and baptized these proofs as ``all-vs-nothing'' proofs. An interesting challenge was finding all-vs-nothing proof but requiring just two observers. A solution was presented in \cite{Cabello01a,Cabello01b}, and combines the magic square with a maximally entangled state between two systems of dimension $4$. This configuration of state and measurements gave raise, on the one hand, to an experimental proposal \cite{CPZBZ03} and a series of experiments \cite{CBPMD05,YZZYZZCP05} testing the all-vs-nothing proof, and, on the other hand, to a variety of nonlocal games stating with \cite{Aravind02}; see Sec.~\ref{sec:3}. See also how is this connected to bipartite Bell inequalities in Sec.~\ref{sec:5}.


\subsection{Magic sets and nonlocal games} 
\label{sec:3}


The adjective ``magic'' was first used by Aravind in Ref.~\cite{Aravind04}. There, Aravind converted the logical demonstration in \cite{Cabello01b} into the following game. Consider two players, Alice and Bob, who cannot communicate during the game. A referee gives input $x \in X$ to Alice and input $y \in Y$ to Bob. Alice returns output $a \in A$ to the referee and, similarly, Bob returns $e \in E$. The referee decides whether the players win or lose based on a winning condition known in advance. A quantum strategy consists of a set of measurements on a Hilbert space ${\cal H}_A$ for Alice, a set of measurements on a Hilbert space ${\cal H}_E$ for Bob, and pairs of systems in an entangled state $|\psi\rangle \in {\cal H}_A \otimes {\cal H}_E$. In a ``magic'' game, each of Alice and Bob are given a context of a magic set and outputs either $+1$ or $-1$ for each variable in the given context. To win, Alice's and Bob's outputs must be the same for any shared variable, and the product of the outputs must be $+1$ for positive contexts and $-1$ for negative contexts. 

One reason why magic games are interesting is because they do not allow for a perfect (i.e., giving winning probability equal to one) classical strategy (due to the parity proof mentioned before), but they allow for a perfect quantum strategy. This is particularly interesting at the light of the observation \cite{JNVWY20} that, in general, it is undecidable to tell whether a nonlocal game has optimal quantum winning probability equal to $1$ or $\le 1/2$, given that one of the two possibilities is the case. 

Magic games belong to a broader family of games called binary constraint system games, which contains examples that do not have a perfect quantum strategy \cite{CM14}. The literature on nonlocal games inspired by magic sets is particularly abundant and starts with the aforementioned paper by Aravind and Ref.~\cite{CHTW04} by Cleve {\em et al.} Nonlocal games inspired by magic sets are also found under the name ``quantum pseudo-telepathy'' \cite{BBT05}.


\subsection{Magic sets and state-independent noncontextuality inequalities}
\label{sec:4}


The proofs of impossibility of hidden variables of Kochen and Specker \cite{KS67}, Peres \cite{Peres90}, and Mermin \cite{Mermin90} rely on assumptions that hold in quantum mechanics for particular quantum systems, but not for general noncontextual hidden-variables theories. In contrast to that, noncontextuality (NC) inequalities hold under the sole assumption of outcome noncontextuality (similarly as Bell inequalities hold under the sole assumption of local realism), without any reference to quantum mechanics. A particularly important class of NC inequalities are the so-called state-independent NC (SI-NC) inequalities, which are violated by any quantum state of any given quantum system (and not only for some entangled states, as Bell inequalities) \cite{Cabello08}. The magic square and the magic pentagram inspired the first SI-NC inequalities \cite{Cabello08}, the first experimental tests \cite{KZG09,ARBC09,MRCL10} of contextuality in nature, and the first SI-NC inequalities that can reveal macroscopic contextuality \cite{Cabello10b}. In fact, as we show in Eq.~(1), there is a natural one-to-one connection between magic sets and SI-NC inequalities. More generally, every proof of the Kochen-Specker theorem can be converted into a SI-NC inequality \cite{BBCP09}.


\subsection{Magic sets and fully nonlocal correlations}
\label{sec:5}


The maximum quantum violation \cite{Tsirelson80} of the Clauser-Horne-Shimony-Holt \cite{CHSH69} does not saturate the violation allowed by the principle of nonsignaling \cite{PR94}. An interesting problem is to determine the simplest (i.e., with the smallest number of settings and outcomes) bipartite Bell inequality in which the maximum quantum violation saturates the nonsignaling bound. This can be detected by computing the local fraction of the correlations \cite{EPR92}. The local fraction measures the fraction that can be described by a local model. Given a matrix of correlations $P(a,b|x,y)$, where $x$ and $y$ are Alice's and Bob's settings, respectively, and $a$ and $b$ are Alice's and Bob's outcomes, respectively, consider all
possible decompositions of the form
\begin{equation}
\label{epr}
P(a,b|x,y) = q_L P_L(a,b|x,y) + (1 - q_L) P_{NL}(a,b|x,y),
\end{equation}
where $P_L(a,b|x,y)$ is a matrix of local correlations and $P_{NL}(a,b|x,y)$ is a matrix of nonlocal nonsignalling correlations. The respective weights,
$q_L$ and $1 - q_L$, satisfy $0 \leq q_L \leq 1$. The local fraction of
$P(a,b|x,y)$ is defined as the maximum of $q_L$ over all
possible decompositions of the form \eqref{epr}.
``Fully nonlocal'' \cite{AGA12} correlations occur if $p_L = 0$ and correspond to the case in which the correlations are as nonlocal as allowed by the principle of nonsingnaling.

The simplest example of bipartite fully nonlocal quantum correlations is the one obtained when Alice measures the rows and Bob measures the columns of the magic square on a maximally entangled state of two ququarts \cite{AGA12}. These correlations maximally violate the two-party, three-setting, four-outcome Bell inequality introduced in \cite{Cabello01b}, based on the magic square, which was proven to be a tight Bell inequality in \cite{GMS07} and was experimentally tested in \cite{CBPMD05,YZZYZZCP05}. With more generality, any magic set can be used to produce an example of bipartite fully nonlocal quantum correlations \cite{AGA12}.


\subsection{Magic sets and proofs with vectors of the Kochen-Specker theorem}
\label{pc}


By applying a method proposed by Peres \cite{Peres91} (see also \cite{KP95,DP97}), any magic set can be converted into a set of rank-one projectors such that, for each set of $d=2^n$ mutually orthogonal rank-one projectors, it is impossible to assign the value $1$ to one projector and the value $0$ to the other $d-1$ projectors, as in the proof of impossibility of hidden variables of Kochen and Specker \cite{KS67}.


\subsection{Magic sets in quantum computation} 


While studying the computational power of correlations used in measurement-based quantum computation \cite{RB01}, Anders and Browne \cite{AB09} found an intriguing relationship between the violation of local realistic models and the computational power of entangled resource states. The connection with magic sets follows from the observation that, e.g., GHZ states (the example used by Anders and Browne) are the only common eigenvectors of the four mutually commuting nonlocal Pauli observables in the magic pentagram, while the local measurements needed to produce (maximal) Bell nonlocality are precisely the other observables in the pentagram. Then, Raussendorf \cite{Raussendorf13} showed that measurement-based quantum computations which compute nonlinear Boolean functions with a high probability (including an example which has a superpolynomial speedup over the best-known classical algorithm, namely, the quantum algorithm that solves the ``discrete log'' problem) are contextual (they violate a noncontextuality inequality). 

For further developments in measurement-based quantum computation in which magic states play a crucial role, see Refs.~\cite{DGBR15,RBDOB17}.

The magic square also appears in the first proof of nonoracular quantum speedup \cite{BGK18}.


\subsection{Magic sets in nonlocality based on local contextuality}


This area comprises several methods for converting contextuality experiments based on sequential measurements on single systems into tests of Bell nonlocality involving pairs of such systems \cite{Cabello10,LHC16,Cabello21}. The magic square inspired one of these methods \cite{Cabello10} and the first experiment of Bell nonlocality based on Kochen-Specker contextuality with sequential measurements \cite{LHC16}.


\subsection{Magic sets in device-independent quantum key distribution} 


Horodecki {\em et al.}~\cite{HHHHPB10} showed that, if Alice and Bob share a magic square (Alice has the rows and Bob the columns) and, in addition, a maximally entangled state of two ququarts, then they can use them to extract secure key in a device-independent manner (i.e., by observing only the input-output statistics of a Bell inequality-like experiment, without making assumptions about the inner functioning of the preparation and measurement devices). 

While standard quantum key distribution (QKD) require a sequential execution of bipartite games, Jain, Miller, and Shi \cite{JMS20} proved the security of a device-independent (DI) QKD protocol based on the magic square where all games are executed in parallel. This result reduces the security requirements for DI-QKD by allowing arbitrary information leakage of each of Alice's and Bob's inputs. The protocol tolerates a constant level of device imprecision and achieves a linear key rate.


\subsection{Magic sets in state-independent quantum dimension witnessing} 


The idea \cite{GBCKL14} is that some forms of quantum contextuality can be used to certify lower bounds on the dimension of the quantum system in experiments with sequential measurements. Interestingly, there is a dimension witness based on the magic square that works independently of the
prepared quantum state and is robust against noise and imperfections, including the case that the measurements are not commuting projective
measurements \cite{GBCKL14}.


\subsection{Magic sets in device-independent self-testing}


Device-independent self-testing (DI-ST) allows for certifying the quantum state and the measurements, up to local isometries, using only the input-output statistics observed. Magic sets are used for DI-ST in \cite{WBMS16,KM17,CS17,CMMN20}. DI-ST is related to rigidity of games. A game is rigid if a near-optimal score guarantees, under the sole assumption of the validity of quantum theory, that the players are using an approximately unique quantum strategy. The magic square allows for self-testing two two-qubit maximally entangled states and the magic square game is rigid \cite{WBMS16}, the magic pentagram game is rigid \cite{KM17}. If the solution group of a magic game is such that there is a unique nontrivial irreducible representation (up to unitary equivalence), then a magic game is rigid \cite{CS17}. However, there are magic games that are not rigid \cite{CMMN20}.


\subsection{Magic sets and the connection between quantum Bell nonlocality and graph invariants}


As explained in Sec.~\ref{pc}, any magic set can be converted into a proof with vectors of the Kochen-Specker theorem \cite{KS67}. In Ref.~\cite{Cabello21} it is shown that it is possible to produce bipartite quantum correlations whose local bound and quantum bounds correspond to different graph invariants of the graph of orthogonality of the Kochen-Specker set of rank-one projectors.


\subsection{Magic sets in quantum gravity}


Magic sets have also been connected to structures that provide a unifying finite geometric underpinning for understanding the structure of functionals used in theories of gravity and black hole entropy \cite{LHS17}.


\section{Methodology}


Here we add details on the methods we have used in the main text for identifying magic sets. This section includes several results from algebraic graph theory and linear algebra. 

A hypergraph $H = (V, E)$ is a finite set $V$ of vertices (in our case, each of them representing a Pauli observable) and a finite set $E$ of hyperedges (in our case, each of them representing a set of mutually commuting Pauli observables), where each hyperedge is a multi-set of vertices. The \emph{multiplicity} of a vertex $v \in V$ and hyperedge $e \in E$ is the number of times $v$ occurs in $e$, and the \emph{degree} of $v\in V$ is the sum of multiplicities of $v$ and $e$ over all hyperedges $e \in E$. 
We define an \emph{Eulerian hypergraph} to be a hypergraph where each vertex has even degree. We also define a \emph{proper Eulerian hypergraph} to be an Eulerian hypergraph with no isolated vertices, no empty hyperedges, no repeated hyperedges, and no repeated vertices in an hyperedge.
Unless stated otherwise, in this appendix we will denote the number of vertices as $|V| = m$ and the number of hyperedges as $|E| = n$.

For a Hilbert space $\mathcal{H}$, we say that an assignment $\alpha : V \to GL(\mathcal{H})$ is \emph{magic} if 
\begin{enumerate}
\item $ \alpha(v)^2 = I$ and $\alpha(v)$ is Hermitian for all $v\in V$.
\item $\alpha(v)\alpha(w) = \alpha(w)\alpha(v)$ whenever $v, w$ are in a common hyperedge $e \in E$.
\item $\prod\limits_{v\in e} \alpha(v) = \pm I$ for each hyperedge $e \in E$.
\item $\prod\limits_{v \in e} \alpha(v) = -I$ for an odd number of hyperedges $e \in E$.
\end{enumerate}
If there exists a magic assignment for $H$, we say that the hypergraph $H$ is \emph{magic}.
The magic assignment $\alpha$ constitutes a magic set, the $\alpha(v)$ are the observables and the multisets $\{\alpha(v) : v \in e\}$ are the contexts.

We consider only magic assignments $\alpha : V \to \mathcal{P}_k$ where $\mathcal{P}_k$ is the $k$-qubit Pauli group for some $k \in \mathbb{Z}_{> 0}$, in which case we say that $\alpha$ is \emph{Pauli-based}. 

We make use of the binary symplectic representation in which a Pauli matrix in $\mathcal{P}_k$
is represented by $2k$-dimensional vector
over $\mathbb{Z}_2$. The four elementary
Pauli matrices are represented as
\begin{equation}
I = (0, 0),\ X = (1, 0),\ Y = (0, 1),\ Z = (1, 1).
\end{equation}
For a \(k\)-qubit Pauli
matrix, the \(i\)th and $(k+i)$th coordinates
of the corresponding binary vector represent
the \(i\)th Pauli in the tensor product. For example, we represent $IXYXZ$ as $(0, 1, 0, 1, 1 | 0, 0, 1, 0, 1)$. Multiplication can be expressed by summing the
corresponding vectors (however the phase is not taken into account) and the commutativity of two \(k\)-qubit Pauli
matrices with symplectic representations $s_1, s_2$ can be checked using the symplectic product defined by
\begin{equation}
\Omega_k(s_1, s_2) = s_1 N_k s_2^T,
\end{equation} 
where \[N_k =
\begin{bmatrix}
0 & I_k \\
I_k & 0 
\end{bmatrix}
\] is a $2k \times 2k$ block matrix,
and $I_k$ is the $k\times k$ identity matrix.
Two Pauli operators commute if the corresponding symplectic product is $0$, and anticommute if it is $1$. If the sum of a set of symplectic vectors with pairwise symplectic product $0$ is $0$, then the product of the corresponding Pauli matrices is $\pm I$.
Note that the symplectic product is 
nondegenerate, bilinear, and symmetric.

Let $\alpha : V \to \mathbb{Z}_2^{2k}$ be a symplectic assignment of the vertices for some $k \in \mathbb{Z}_{\geq 0}$.
The \emph{Gram matrix} of $\alpha$ is the $m \times m$ matrix $M$
such that
\begin{equation}
M_{i,j} = \Omega_k(\alpha(v_i), \alpha(v_j))
\end{equation}
for $1 \leq i, j \leq m$. The Gram matrix expresses whether any pair of observables commutes or anticommutes. Thus, all the diagonal entries of $M$ are $0$, and it is symmetric. There can be many different assignments which have the same
Gram matrix $M$. We say that such assignments \emph{respect} $M$.

We will show that if one is given a proper Eulerian hypergraph $H$ and a
Gram matrix $M$, this yields enough information to check whether any
Pauli-based assignment $\alpha$ respecting $M$ is magic or not. In other words,
we do not need to search for assignments, but for Gram matrices.

We call an assignment \(\alpha : V \to \mathbb{Z}_2^{2k}\) \emph{valid} if $\Omega_k(\alpha(v_i), \alpha(v_j)) = 0$ whenever $v_i, v_j$ occur in a common context, and if $\sum_{v \in e} \alpha(v) = 0$ for all $e \in E$. Note that the corresponding $k$-qubit Pauli-based assignment satisfies conditions 1--3 of being a magic assignment.
Similarly, if a Gram matrix $M$ satisfies:
\begin{enumerate}
\item $M_{i,j} = 0$ whenever vertices $v_i,v_j$ occur in the same context,
\item $\sum\limits_{v_i \in e} M_{i,j} = 0$, for all $1 \leq j \leq m$ for all contexts $e \in E$,
\end{enumerate}
then we say that $M$ is a \emph{valid Gram matrix for $H$}. 
\begin{lem} 
Let $s_1, \dots, s_m \in \mathbb{Z}_2^{2k}$ for some positive integer $k$, and let $M$ be the $m \times m$ Gram matrix associated to the symplectic product $\Omega_k$, with rows $r_1, \dots, r_m$. If a subset $\{r_{i_1}, \dots, r_{i_j}\}$ of the rows is linearly independent, then the corresponding set of vectors $\{s_{i_1}, \dots, s_{i_j}\}$ is also linearly independent.
\end{lem} 
\proof{For simplicity of notation, assume that the subset is $\{s_1, \dots, s_j\}$. Then,
\begin{align*}
&c_1s_1 + \dots + c_js_j = 0 \\
\implies & \Omega_k(c_1s_1 + \dots + c_js_j, s_l) = 0 \text{ for } 1 \leq l \leq m \\
\implies &c_1\Omega_k(s_1, s_l) + \dots + c_j\Omega_k(s_j, s_l) = 0 \text{ for } 1 \leq l \leq m \\
\implies 
&\begin{cases}
c_1M_{1,1} + \dots + c_jM_{j,1} = 0 \\
\hspace{0.6cm} \vdots \hspace{2.1cm} \vdots \\
c_1M_{1,m} + \dots + c_jM_{j,m} = 0
\end{cases} \\
\implies & c_1r_1 + \dots + c_jr_j = 0.
\end{align*}
}

\begin{prop}
Let $\alpha : V \to \mathbb{Z}_2^{2k}$ be a valid assignment, and let $M$ be the corresponding Gram matrix. Then, $M$ is a valid Gram matrix.
\end{prop}

\proof{This is immediate from the previous Lemma.}

\begin{prop}
Let $M$ be a valid Gram matrix of proper Eulerian hypergraph $H = (V, E)$ with binary rank $2k$ for some $k \in \mathbb{Z}_{\geq 0}$. Then any assignment $\alpha : V \to \mathbb{Z}_2^{2k}$ respecting $M$ is a valid assignment of $H$.
\end{prop}

\proof{Observables in the same context are represented by commuting operators since $M_{i,j} = 0$ whenever $v_i, v_j \in e$ for some $e \in E$. 
We show that, for any $S \subseteq \{1, \dots, m\}$, if $\sum_{i \in S} r_i = 0$, then $\sum_{i \in S} \alpha(v_i) = 0$.
\\ \indent
First note that since the binary rank of $M$ is $2k$, we may assume 
without loss of generality
that rows $r_1, \dots, r_{2k}$ form a basis for $\operatorname{row}(M)$,
the row space of $M$.
By the previous Lemma, it follows that $\alpha(v_1), \dots, \alpha(v_{2k})$ form a basis for $\mathbb{Z}_2^{2k}$. Then,
\begin{align*}
&\sum_{i \in S} r_i = 0 \\
\implies & \sum_{i \in S} M_{i,1} = 0, \dots, \sum_{i \in S} M_{i,m} = 0 \\
\implies &\sum_{i \in S} \Omega_k(\alpha(v_i), \alpha(v_1)) = 0, \dots, \sum_{i \in S} \Omega_k(\alpha(v_i), \alpha(v_m)) = 0 \\
\implies &\Omega_k(\sum_{i \in S} \alpha(v_i), \alpha(v_1)) = 0, \dots, \Omega_k( \sum_{i \in S} \alpha(v_i), \alpha(v_{2k})) = 0 \\
\implies &\Omega_k(\sum_{i \in S} \alpha(v_i), s) = 0 \text{ for all } s \in \mathbb{Z}_2^{2k} \\
\implies &\sum_{i \in S} \alpha(v_i) = 0
\end{align*}
and so we see that the product of observables in any context is $\pm I$ since the sum of the corresponding symplectic vectors is $0$ and each of the vector pairs have symplectic product $0$.}

Note that the previous proposition only applies to $k$-qubit assignments -- there may be assignments $\alpha : V \to \mathbb{Z}_2^{2l}$ (for $l > k$) respecting $M$ which are not valid assignments.

\begin{prop} 
The set of valid Gram matrices of $H$ forms a subspace of the $m \times m$ binary matrices.
\end{prop}

\proof{The entries of $M$ are defined by a set of homogeneous linear equations.}

We call the subspace from the previous proposition the \emph{valid Gram space of \(H\)}. In order to compute the valid Gram space of $H$, we simply have to solve the appropriate set of at most $m^2 + mn$ linear equations. 

\begin{theo} Fix an arbitrary ordering of all contexts and list
them all jointly as 
L=\((L_1,\ldots, L_t)\), where each \(L_i\) is a vertex; the ordering of vertices inside each hyperedge is also arbitrary. Impose a total order $\prec$ on the set of vertices. Let $M$ be a valid Gram matrix, let $\alpha : V \to \mathbb{Z}_2^{2k}$ be a valid assignment respecting $M$ for some $k \in \mathbb{Z}_{\geq 0}$, and let $\overline{\alpha} : V \to \mathcal{P}_k$ be the corresponding $k$-qubit assignment.
Then $\overline{\alpha}$ is a magic assignment if and only if
\begin{equation}
\sum\limits_{1\le i < j\le t,\ L_i \succ L_j} 
M_{L_i,L_j} = 1.
\end{equation}
\end{theo}

\proof{Consider the product
\[
\prod_{i=1}^t \overline{\alpha}(L_i) = 
\prod_{e \in E} \prod_{v \in e} \overline{\alpha}(v).
\]
After swapping pairs of 
adjacent operators
whenever necessary
(which corresponds to performing the bubble sort
with respect to the order $\prec$ on the list $L$), we can rewrite this product in the form $(-1)^s \prod_{v \in V} \overline{\alpha}(v)^{deg(v)}$, where 
$s = \sum\limits_{1\le i < j\le t,\ L_i \succ L_j}
M_{L_i,L_j} $. Since each operator squares to $I$, the product simplifies to $(-1)^s I$ and the result follows.
}

Thus, if $M$ is a valid Gram matrix for $H$, then it is easy to check whether or not there exists a Pauli-based proof satisfying (i) and (ii) respecting $M$ or
not. If it does exist, then we call this Gram matrix \emph{magic}.

Given $k \in \mathbb{Z}_{\geq 0}$, the \emph{symplectic graph SP(2k)}
is the graph whose vertices are the $2^{2k} - 1$ non-zero
binary vectors of length $2k$, and an edge exists between two vertices if 
and only if
the corresponding vectors have symplectic product 1. A graph is called \emph{reduced} if it has no isolated vertices and no
pair of vertices have the same neighbourhoods. A graph $G'$ is an
\emph{induced subgraph} of a graph $G$ if $G'$ can be obtained from
$G$ by a sequence of vertex deletions.

\begin{theo}
\label{th-GR}\cite[Theorem 8.11.1]{GR01} If a graph $G$ is reduced and its adjacency matrix
has binary rank 
at most
$2k$ for some $k \in \mathbb{Z}_{\geq 0}$, then $G$ is an induced subgraph of $SP(2k)$.
\end{theo}

From the $m \times m$ Gram matrix $M$ we can construct the graph $G = (V, E)$ such that $M$ is the adjacency matrix of $G$ by taking the vertices to be $V := \{1, \dots, m\}$ and the edges to be the pairs $\{i,j\}$ such that $M_{i,j} = 1$. If $G$ is reduced and $M$ has binary rank $2k$, then we can find an isomorphic copy of $G$ as an induced subgraph of $SP(2k)$ by Theorem~\ref{th-GR}. This yields a valid assignment $\alpha : V \to \mathbb{Z}_2^{2k}$ of $H$, which yields a parity proof $\overline{\alpha} : V \to \mathcal{P}_k$ if $M$ is a magic Gram matrix. Computing an isomorphic copy of $G$ in $SP(2k)$ can be done in a straightforward manner in \emph{SageMath} for example, and one can also iterate through all such copies. 

One can improve the approach in the previous paragraph. Let $B$ denote a set of row indices of $M$ whose corresponding rows form a basis for $\operatorname{row}(M)$. Due to the bilinearity of the symplectic product $\Omega_k$, one only needs to find a partial assignment $\alpha'$ valid for the vertices $\{v_i : i \in B\}$. Then, for any given vertex $v_j \in V$, we have
\[
r_j = \sum_{i \in B} a_ir_i
\]
for some $a_i \in \mathbb{Z}_2$, and so the assignment $\alpha : V \to \mathbb{Z}_2^{2k}$ given by
\[
\alpha(v_j) := \sum_{i \in B} \alpha'(v_i)
\]
for all $v_j \in V$ is a magic assignment. In order to compute the partial assignment $\alpha'$, we define $M_B$ to be the submatrix of $M$ by taking only rows and columns whose row (respectively column) index lies in $B$, and we define $G_B$ to be the Gram matrix corresponding to $M_B$. Then $\alpha'$ can be obtained by finding an isomorphic copy of $G_B$ in $SP(2k)$. Finally, we remark that in the case when $SP(2k)$ is too large, one can simply apply a backtracking procedure to generate the partial assignment $\alpha'$ (this was the approach taken to generate a 6-qubit magic assignment for MS6-35, see Appendix~\ref{ap:6-qubit}).

In the case that $G$ is not reduced we can still construct an assignment using operators of $\mathcal{P}_k$. We first observe that if $G$ is not reduced, this can be identified directly in $M$. Namely, if $G$ contains an isolated vertex, then the corresponding row of $M$ is all zeroes, and if vertices of $G$ have the same neighbourhood, then the corresponding rows of $M$ are identical. We thus say that a Gram matrix $M$ is \emph{reduced} if $M$ has no zero-row or repeated rows. One can thus \lq reduce\rq\ $M$ by removing all rows (and corresponding columns) of zeroes, and replacing sets of identical rows (and their corresponding columns) by a single copy to create a new Gram matrix $M'$ which is reduced and has the same binary rank as $M$. Once an assignment is found satisfying the vertices corresponding to $M'$, it can be extended to a assignment of $H$ by assigning the identity operator to all zero rows, and assigning the same operator to identical rows. 

\begin{theo} 
\label{thm-magic-test-appendix}
A proper Eulerian hypergraph $H$ has a magic assignment with Pauli observables for a system of $k$ qubits satisfying (i) and (ii) if and only if there is a magic Gram matrix of binary rank at most $2k$ in the valid Gram space of $H$.
\end{theo}

We note a subtle point. If $H$ has no magic Gram matrix in its valid Gram space, then $H$ admits no Pauli-based magic assignment satisfying (i) and (ii), but it may still have a magic assignment satisfying (i) and (ii) using operators from $GL(\mathcal{H})$ for some Hilbert space $\mathcal{H}$. We know of no such example where this occurs.

\begin{cor} For a proper Eulerian hypergraph $H$, the minimum $k$ such that $H$ has a $k$-qubit parity proof is half of the minimum binary rank over all magic Gram matrices in the valid Gram space of $H$.
\end{cor}

For a proper Eulerian hypergraph $H$ admitting a Pauli-based magic assignment satisfying (i) and (ii), the magic Gram matrices also have a special structure within the valid Gram space of $H$, namely, they form an affine space of exactly half the size of the valid Gram space. This is straightforward to show since the sum of two magic Gram matrices is nonmagic, the sum of two non-magic Gram matrices is also nonmagic and the sum of a pair of nonmagic and magic Gram matrices is magic. Thus we also see that the set of nonmagic Gram matrices forms a subspace of the valid Gram space. Therefore, in order to prove that a proper Eulerian hypergraph $H$ is magic, we simply need to check whether or not there is a magic Gram matrix in a basis for the valid Gram space.

This concludes the proof of Theorem~1 from the main text.

We define a proper Eulerian hypergraph $H$ to be \emph{minimal} if for any Pauli-based magic assignment of $H$ satisfying (i) and (ii) no pair of vertices is mapped to the same operator, and no vertex is mapped to the identity operator. 

\begin{theo} \label{theo:minimal} Let $H$ be a proper Eulerian hypergraph. Then $H$ is minimal if and only if every magic Gram matrix $M$ in the valid Gram space of $H$ is reduced.
\end{theo}

\proof{If there is some Gram matrix $M$ in the valid Gram space of $H$ which is not reduced, then there is a Pauli-based magic assignment of $H$ satisfying (i) and (ii) which utilizes either the identity operator or has two vertices assigned to the same operator. Conversely, any valid assignment $\alpha$ corresponding to a reduced Gram matrix $M$ in the valid Gram space of $H$ has no $\pm I$ operator and no pair of vertices assigned to the same operator.
}

In the case that a proper Eulerian hypergraph $H$ is not minimal, it can be used to obtain minimal proper Eulerian hypergraphs using the following procedure:
\begin{enumerate}
\item Choose a reducible magic Gram matrix $M$ in the valid Gram space of $H$.
\item For each row $i$ of zeroes of $M$, the corresponding vertex $v_i$ is deleted. 
\item If rows $i_1,\ldots, i_j$ of $M$ are identical, vertices $v_{i_1},\ldots, v_{i_j}$ are identified (i.e., each vertex is relabeled $v_{i_1}$)
\item After the identification, we may get repeated vertices in a given hyperedge. In this case, we reduce the multiplicity of each vertex, hyperedge pair modulo $2$.
\item At this stage we may have repeated hyperedges. In this case, we reduce the number of occurrences of each hyperedges modulo $2$. We then delete any empty hyperedges.
\item We now obtain a new proper Eulerian hypergraph $J$. If $J$ is minimal, we return it. If not, we recurse on $J$.
\end{enumerate}
The hypergraph $J$ obtained in Step $6$ is necessarily magic - it is straightforward to check that if $M'$ is the matrix obtained by reducing $M$, then $M'$ is a magic Gram matrix for $J$. Note that only assignments of $J$ respecting $M'$ extend to assignments of $H$, and thus for a minimal Eulerian hypergraph $K$ found at a recursion depth $\geq 2$, it may be possible that no assignment of $K$ extends to an assignment of $H$.

Combining our results with Arkhipov's theorem yields a novel algorithm for determining the planarity of a graph. Let $H = (V, E)$ be a proper hypergraph with $V = \{v_1, \ldots, v_m\}$ and $E = \{e_1, \ldots, e_n\}$. The \emph{dual} of $H$ is the hypergraph $H^* = (X, F)$ with $X = \{x_1, \ldots, x_n\}$ and $F = \{f_1, \ldots, f_m\}$ such that the multiplicity of $x_i$ and $f_j$ is equal to the multiplicty of $v_j$ and $e_i$ for any pair $1 \leq i \leq m$, $1 \leq j \leq n$. For the following result, note that a simple graph is implicitly a proper hypergraph, so that we may refer to the dual hypergraph of a simple graph.

\begin{cor} \label{cor:graphs}
Let $G$ be a simple graph, and let $H$ be the dual hypergraph of $G$. Let $\mathcal{B}$ be a basis for the valid Gram space of $H$. Then, $G$ is nonplanar if and only if there is a magic Gram matrix in $\mathcal{B}$.
\end{cor}

We also remark that there will be always be a magic gram matrix $M$ in the valid Gram space of $H$ such that $M$ encodes the exact operations needed to find a topological $K_5$ or $K_{3,3}$ minor in the case that $G$ is nonplanar. Our result draws a new connection between algebraic and topological graph theory.


\section{Noncontextuality inequalities based on magic sets}


Here we add details to the section on noncontextuality inequalities in the main text.


\subsection{Proof of Theorem~2}


Let $H = (V, E)$ be a magic proper Eulerian hypergraph with $V = \{v_1, \dots, v_m\},\ E = \{e_1, \dots, e_n\}$. For a magic assignment $\alpha$ of $H$ and $i=1,\dots,n$, we let
\[
sgn_{\alpha}(e_i) := 
\begin{cases}
1 &\text{ if } \prod_{v \in e_i} \alpha(v) = I \\
-1 &\text{ if } \prod_{v \in e_i} \alpha(v) = -I
\end{cases}
\]
and
\[
c_i := 
\begin{cases}
0 &\text{ if } sgn_{\alpha}(e_i) = 1 \\
1 &\text{ if } sgn_{\alpha}(e_i) = -1
\end{cases}
\]
with $c(\alpha) = (c_1, \dots, c_n)$. We follow the same convention for classical assignments, with the slight modification $sgn_a(e_i) = \prod_{v \in e_i} a(v)$. Let $\alpha$ be a magic assignment of $H$. The \em noncontextual bound for $\alpha$ \em is given by 
\begin{equation}
b_{\alpha}(H) := \max_{\text{classical assignment } a} \sum\limits_{e \in E} sgn_\alpha(e) sgn_a(e).
\end{equation}
We remark that this is $b$ from the main text - we use this notation to emphasize the dependence of the noncontextual bound on the hypergraph $H$ and the magic assignment $\alpha$.

Recall that for a hypergraph $H = (V, E)$ with $|V| = m,\ |E| = n$, the matrix $M \in \mathbb{Z}_2^{m \times n}$ defined by
\[
M_{i,j} = 
\begin{cases}
1 \text{ if } v_i \in e_j \\
0 \text{ if } v_i \notin e_j
\end{cases}
\]
is called the \em incidence matrix \em of $H$. Let $\mathcal{C}$ denote the set of all possible $2^n$ classical assignments $a : V \to \pm 1$.

\begin{prop} \label{prop:row-space}
Let $H = (V, E)$ be a proper Eulerian hypergraph with incidence matrix $M$. 
Define $\phi : \mathbb{Z}_2^m \to \mathcal{C}$ via $\phi(x) = a$, where $a(v_i) = (-1)^{x_i}$ for all $i=1,\dots,m$. Then $\phi$ is a bijection, and $x^TM = c(a)$ whenever $\phi(x) = a$.
\end{prop}

\proof{Clearly, $\phi$ is a bijection.
Let $x \in \mathbb{Z}_2^m$, $y = x^TM$, and $d_1, \dots, d_n$ denote the columns of $M$ corresponding to hyperedges $e_1, \dots, e_n$ respectively. Then for any $1 \leq j \leq n$, we have $y_j = x \cdot d_j$, so $y_j = 1 \iff sgn_a(e_j) = -1$. Thus we see that $y = c(a)$, and the result follows.
}

The previous result shows that for any classical assignment $a : V \to \pm 1$ of $H$, the vector $c(a)$ encoding the hyperedge products $\prod_{v \in e} a(v)$ is an element of the rowspace of the incidence matrix $M$, and vice-versa.

\begin{prop}
Let $H = (V, E)$ be a magic proper Eulerian hypergraph with incidence matrix $M$. Let $\alpha$ and $\alpha'$ be two magic assignments of $H$. If $c(\alpha) - c(\alpha') \in \operatorname{row}(M)$, then $b_\alpha(H) = b_{\alpha'}(H)$.
\end{prop}

\proof{Let $a : V \to \pm 1$ be a maximizer of the sum
\[
\sum_{e \in E} sgn_{\alpha}(e)sgn_a(e).
\]
By Proposition \ref{prop:row-space} there exists some classical assignment $a' : V \to \pm 1$ such that $c(a') = c(\alpha) - c(\alpha')$, so for any $e \in E$, we have $sgn_{\alpha}(e) = sgn_{\alpha'}(e)sgn_{a'}(e)$. Then $\sum_{e \in E} sgn_{\alpha}(e)sgn_{a}(e) = \sum_{e \in E} sgn_{\alpha'}(e)sgn_{a''}(e)$ where $a''(v) = a(v)a'(v)$ for all $v \in V$. Thus $b_{\alpha}(H) \leq b_{\alpha'}(H)$. Since we are working over $\mathbb{Z}_2$, $c(\alpha) - c(\alpha') = c(\alpha') - c(\alpha)$ so the same argument works in the opposite direction.
The result follows.
}

Therefore, for a given magic assignment $\alpha$, for any magic assignment $\alpha'$ with $c({\alpha'})$ in the affine space $c(\alpha) + \operatorname{row}(M)$, we have $b_\alpha(H) = b_{\alpha'}(H)$. 

Viewing $\operatorname{row}(M)$ as a linear code $C$, we use results from coding theory in order to compute the noncontextual bound for a given magic assignment $\alpha$. Define $C'$ to be the space generated by $C \cup \{ c(\alpha) \} $. Then a smallest weight element of the coset $\operatorname{row}(M) + c(\alpha)$ is a smallest odd weight element of $C'$.
The smallest odd weight occurring in $C'$ can
be computed either directly, or using
the dual of $C'$ and MacWilliams identity.

By examining an element $y^* \in \mathbb{Z}_2^n$ of lowest Hamming weight (denote Hamming weight of a vector $y$ by $w(y)$) in $c(\alpha) + \operatorname{row}(M)$, we get our main result. The key observation is that the assignment $a$ setting each vertex to $1$ will yield the noncontextual bound for any magic assignment $\alpha^*$ with $c(\alpha^*) = y^*$. 

We now prove Theorem~2. 

\proof{Let $y^*$ be a lowest Hamming weight element of the affine space $c(\alpha) + \operatorname{row}(M)$.
We first show that $b_{\alpha}(H) \geq |E| - 2w(y^*)$. Since $y^* \in c(\alpha) + \operatorname{row}(M)$, there exists $x \in \mathbb{Z}_2^n$ such that $y^* = c(\alpha) + x^TM$. Define $a : V \to \pm 1$ by $a(v_i) = (-1)^{x_i}$ for $i=1,\dots,m$. Then, $\sum_{e \in E} sgn_{\alpha}(e)sgn_a(e) = |E| - 2w(y^*) \leq b_{\alpha}(H)$. 
\\ \indent
Now assume towards a contradiction that $b_{\alpha}(H) > |E| - 2w(y^*)$. Then there exists $a : V \to \pm 1$ such that $\sum_{e \in E} sgn_{\alpha}(e)sgn_a(e) > |E| - 2w(y^*)$. Let $y' = c(\alpha) + c(a)$. Then clearly, $\sum_{e \in E} sgn_{\alpha}(e)sgn_a(e) = |E| - 2w(y')$, and so $w(y') < w(y^*)$. However, since $a$ is a classic assignment, by Proposition~\ref{prop:row-space}, we have $c(a) \in \operatorname{row}(M)$, so $y' \in c(\alpha) + \operatorname{row}(M)$ contradicting the minimality of $y^*$.
The result follows.
}

Note that while a magic assignment $\alpha^*$ with $c(\alpha^*) = y^*$ is guaranteed to exist for NCHV models (we simply have to negate the proper operators), its existence is immaterial to the proof of Theorem~2. In particular, this theorem can be used to compute noncontextual bounds for Pauli-based magic assignments.

We remark that one can also compute $b_{\alpha}(H)$ by iterating over all $2^m$ possible classical assignments $a$. The most costly step in our method is finding the lowest odd weight in $C'$ (or its dual), which takes roughly $2^{\min (\dim(C'), {\rm codim}(C'))}$ steps. This value is bounded above by $2^{\lfloor \frac{n+1}{2} \rfloor}$. Thus, for example if the number of observables and contexts are the same, in the worst case, asymptotically the time taken by this algorithm is the square root of the time taken by the naive approach.

The \emph{noncontextual bound of $H$}, $b(H)$ is given by the minimum value of $b_{\alpha}(H)$ over all magic assignments $\alpha$ of $H$. For a coset $C$ of $\operatorname{row}(M)$, define $w(C)$ to be the lowest Hamming weight over all elements of $C$ [equivalently, the smallest number of negative contexts over all magic assignments $\alpha$ with $c(\alpha) \in C$]. The following theorem follows directly from Theorem~2 of the main text.

\begin{theo} \label{theo:ncb}
Let $H = (V, E)$ be a proper Eulerian hypergraph with incidence matrix $M$, and let $\Gamma := \{c(\alpha) + \operatorname{row}(M) : \alpha \text{ is a magic assignment of } H\}$ be the set of cosets of $\operatorname{row}(M)$ with a magic assignment. Let $\mathcal{N} = \max_{C \in \Gamma} w(C)$. Then $b(H) = |E| - 2\mathcal{N}$. 
\end{theo}

We also define the \emph{Pauli-based noncontextual bound of $H$} to be the minimum value of $b_{\alpha}(H)$ over all Pauli-based magic assignments $\alpha$ of $H$. For Pauli-based assignments $\alpha$, $\alpha'$ of $H$, we define the \emph{tensor assignment} $\beta$, where $\beta(v) := \alpha(v) \otimes \alpha'(v)$ for all $v \in V$. Notice that $c(\beta) = c(\alpha) + c(\alpha')$.
For vertices $u,v \in V$, $\beta(u)$ and $\beta(v)$ commute if and only if the pairs $\alpha(u)$, $\alpha(v)$ and $\alpha'(u)$, $\alpha'(v)$ both commute or if they both anti-commute. Thus, if $\alpha$ and $\alpha'$ are magic assignments which respect the same magic Gram matrix $N$, then $\beta$ respects the Gram matrix $N + N = 0$; that is, $\beta$ is a commutative assignment. Therefore, $c(\beta) \in \operatorname{row}(M)$. If all observables commute, then no quantum advantage is gained; see, e.g., the proof of \cite{Arkhipov12} (Proposition~8). Since, $c(\alpha) + c(\alpha') = c(\alpha) - c(\alpha') \in \operatorname{row}(M)$, it follows that $b_{\alpha}(H) = b_{\alpha'}(H)$. In words, Pauli-based magic assignments respecting the same Gram matrix have the same noncontextual bound. Therefore, in Theorem \ref{theo:ncb}, the number of cosets we need to check is at most the number of magic Gram matrices in the valid Gram space of $H$. 
The hypergraph $H$ of MS4-21 in Fig.~2(a) of the main text has a single magic Gram matrix. We are thus able to compute the Pauli-based noncontextual bound of $H$ by computing $b_{\alpha}(H)$ for a single Pauli-based magic assignment $\alpha$. Since the bound for MS4-21 is $15$, we immediately conclude that the Pauli-based noncontextual bound of $H$ is $15$, and by Theorem 2 of the main text that $H$ has no Pauli-based magic assignment with a single negative context.


\subsection{Robustness against imperfections of the violation of noncontextuality inequalities. Tolerated error per context}


Consider the noncontextuality inequalities given by Eq.~(1).
\begin{equation}
\label{nci2}
\chi = \sum_{{\cal C}_i \in C_p} \langle {\cal C}_i \rangle - \sum_{{\cal C}_j \in C_n} \langle {\cal C}_j \rangle \le b.
\end{equation}

Quantum mechanics predicts that, for any initial state, $\langle {\cal C}_i \rangle=1$, if ${\cal C}_i \in C_p$ (the set of positive contexts), and $\langle {\cal C}_j \rangle=-1$, if ${\cal C}_j \in C_n$ (the set of negative contexts). Therefore,
$\chi_{\rm QM} = N$. However, in actual experiments (e.g., \cite{KZG09,
ARBC09, MRCL10}), the experimental values are
$\langle {\cal C}_i^{(m)} \rangle=1-\epsilon_i$ and $\langle {\cal C'}_j^{(m)} \rangle=-1+\epsilon_j$, $0 < \epsilon_i \ll 2$. Therefore, the experimental value is $\chi_{\rm expt}=1- \sum_{k=1}^N \epsilon_k$. The experiments do not reach $\chi_{\rm QM}$ for different reasons, for example, nonperfect unitary operations and entangling gates \cite{KZG09}, and nonperfect
alignment of the interferometric setups \cite{ARBC09}.

Experimental imperfections can also be interpreted as a failure
of the assumption of perfect sharpness (ideality) of the measurements and compatibility of the targeted observables under which the
bound $b$ is valid, and force us to correct this bound. This
correction is of form $b \rightarrow b'= b + \sum_{k=1}^N
\phi_k$, where $\phi_k > 0$ can be obtained from additional experiments \cite{KZG09,GKCLKZGR10,NDSC12} or from the experimentally observed deviation of the perfect nondisturbance \cite{DKL15}.

Assuming that all correlations are affected by similar errors,
i.e., that $\sum_{k=1}^N \epsilon_k = N \epsilon$ and
$\sum_{k=1}^N \phi_k = N \phi$, we can define the error per
correlation as $\varepsilon= \epsilon+\phi$. A natural measure
of robustness of a quantum violation of inequality~(1) against imperfections is the {\em tolerated error
per context}, which can be expressed as
\begin{equation}
\varepsilon = \frac{\chi_{\rm QM}-b}{N}.
\end{equation}
If $\phi$ is negligible, then $\varepsilon$ is the maximum
difference that can be tolerated (still violating the
inequality) between the experimental value of a correlation and
the quantum value for an ideal experiment. There, for a given $n$, higher $\varepsilon$ means higher resistance against imperfections of the violation of the corresponding noncontextuality inequality.

If $\phi$ is not negligible but is similar for experiments with
sequential measurements of the same length, then $\varepsilon$
is a good measure to compare the resistance to imperfections of
inequalities involving correlations between the same number of
measurements. For a different argument supporting this statement
see \cite{KGPLC11}.


\section{Other magic sets found in this work}


\begin{table}[t!]
\label{table1}
\begin{center}
\begin{tabular}{c c c c c l}
\hline \hline
$n$ & Magic set & Observables & Contexts & $b/Q$ & $\varepsilon$ \\
\hline
$2$ & Square & $9$ & $6$ & $4/6$ & $0.33$ \\
$3$ & Pentagram & $10$ & $5$ & $3/5$ & $0.4$ \\
$3$ & MS3-27 & $27$ & $27$ & $21/27$ & $0.22$ \\
$4$ & MS4-21 & $21$ & $21$ & $15/21$ & $0.28$ \\
$4$ & MS4-27 & $27$ & $27$ & $17/27$ & $0.37$ \\
$5$ & MS5-27 & $27$ & $27$ & $17/27$ & $0.37$ \\
\hline
\hline
\end{tabular}
\caption{\label{tab:I}Vertex-transitive irreducible magic sets after this work. $n$ is the number of qubits. $b$ is the bound of the noncontextuality inequality~(1) and $Q$ is the corresponding quantum value. $\varepsilon$ is the error per context that can be tolerated while still violating the noncontextuality inequality.}
\end{center}
\end{table}


\begin{table}[t!]
\label{table1}
\begin{center}
\begin{tabular}{c l c c c l}
\hline \hline
$n$ & Magic set & Observables & Contexts & $b/Q$ & $\varepsilon$ \\
\hline
$3$ & MS3-15 & $10_4 + 5_6$ & $10_3+10_4$ & $14/20$ & $0.3$ \\
$3$ & MS3-18 & $3_2 + 15_4$ & $6_3 + 12_4$ & $12/18$ & $0.33$ \\
$3$ & MS3-27b & $27_4$ & $27_4$ & $17/27$ & $0.37$ \\
$3$ & MS3-29 & $27_4 + 2_{12}$ & $33_4$ & $19/33$ & $0.424$ \\
$4$ & MS4-20 & $5_2 + 15_4$ & $6_3 + 13_4$ & $17/19$ & $0.105$ \\
$4$ & MS4-21b & $11_2 + 10_4$ & $2_3 + 14_4$ & $14/16$ & $0.125$ \\
$4$ & MS4-21c & $1_2 + 19_4 + 1_6$ & $21_4$ & $19/21$ & $0.095$ \\
$4$ & MS4-24 & $3_2 + 11_4 + 9_6 + 1_{10}$ & $2_3 + 12_4 + 12_5$ & $20/26$ & $0.23$ \\
$5$ & MS5-26 & $25_4 + 1_{10}$ & $10_3 + 20_4$ & $24/30$ & $0.2$ \\
$5$ & MS5-29 & $23_4 + 5_6 + 1_8$ & $6_3 + 28_4$ & $28/34$ & $0.176$ \\
$5$ & MS5-31 & $3_2 + 23_4 + 2_6 + 3_8$ & $2_3 + 12_4 + 16_5$ & $24/30$ & $0.2$ \\
$6$ & MS6-35 & $30_4 + 5_8$ & $3_3 + 14_4 + 19_5$ & $30/36$ & $0.167$ \\
\hline
\hline
\end{tabular}
\caption{\label{tab:II}Some non-vertex-transitive irreducible magic sets. $n$ is the number of qubits. $b$ is the bound of the noncontextuality inequality~(1) and $Q$ is the corresponding quantum value. $\varepsilon$ is the error per context that can be tolerated while still violating the noncontextuality inequality. In order to indicate the number of observables of each degree we write $x_y$ to mean that there are $x$ observables appearing in exactly $y$ contexts. Similarly, to indicate the number of observables in each context, we write $x_y$ to mean that there are $x$ contexts containing exactly $y$ observables. 
For example, for MS3-15, $10_4 + 5_6$ in the Observables column indicates that MS3-15 has 10 observables appearing in exactly 4 contexts and 5 observables appearing in exactly 6 contexts. The $10_3+10_4$ in the Contexts column indicates that MS3-15 has $10$ contexts of size $3$ and $10$ contexts of size~$4$.}
\end{center}
\end{table}


Using the algorithm described in the main text, we have found seven magic Eulerian hypergraphs, which we call HA, MS3-27, MS4-27, MS5-27, HB, HC, and HD. 
Three of them, MS3-27, MS4-27, and MS5-27 are minimal, so each of them corresponds to a class of equivalence of irreducible magic sets. These classes have been described in the main text. Some of their properties are collected in Table~\ref{tab:I}. 

We have used the other four hypergraphs to find new minimal Eulerian hypergraphs using the reduction process outlined in the discussion following Theorem~\ref{theo:minimal}. For two of the hypergraphs, HB and HD, our code terminated, and for the other two, HA and HC, we terminated the code after several months as there are $2^{30}$ and $2^{26}$ respectively magic Gram matrices at the first level of recursion. 
For hypergraph HA we have found, so far, $84$ magic sets. From hypergraph HB we have found a total of $309$ magic sets. From hypergraph HC, we have found, so far, $7368$ magic sets. From hypergraph HD we have found a single magic set, MS3-27b. 

In the following, we present assignments requiring a minimal number of qubits for MS3-29, and for some notable magic sets found from HA, HB, HC, and HD. Those with minimal number of observables, or contexts, or maximal resistance to noise (among those found from that hypergraph). We also include straight line representations in the Euclidean plane whenever we have been able to obtain them. Some properties of these non-vertex-transitive minimal sets can be found in Table~\ref{tab:II}.

\subsection{From HA: MS6-35} \label{ap:6-qubit}

The hypergraph HA has the following structure with $45$ observables and $45$ contexts: \\

[[1, 4, 14, 20, 42], [1, 6, 15, 21, 41], [1, 25, 31, 35, 44], [1, 27, 32, 36, 43], [2, 4, 13, 19, 41], [2, 5, 15, 20, 40], [2, 26, 33, 36, 44], [2, 27, 31, 34, 45], [3, 5, 13, 21, 42], [3, 6, 14, 19, 40], [3, 25, 33, 34, 43], [3, 26, 32, 35, 45], [4, 7, 22, 28, 35], [4, 9, 23, 29, 34], [5, 8, 24, 29, 35], [5, 9, 22, 30, 36], [6, 7, 24, 30, 34], [6, 8, 23, 28, 36], [7, 10, 16, 40, 43], [7, 12, 17, 42, 44], [8, 11, 16, 42, 45], [8, 12, 18, 41, 43], [9, 10, 17, 41, 45], [9, 11, 18, 40, 44], [10, 13, 23, 31, 37], [10, 14, 24, 33, 38], [11, 13, 24, 32, 39], [11, 15, 22, 33, 37], [12, 14, 22, 31, 39], [12, 15, 23, 32, 38], [16, 19, 25, 30, 38], [16, 21, 26, 29, 39], [17, 19, 27, 29, 37], [17, 20, 25, 28, 39], [18, 20, 26, 30, 37], [18, 21, 27, 28, 38]]. \\

One of the magic sets obtained from HA, the one called MS6-35 has the following structure with $35$ observables and $36$ contexts:\\

[[1, 3, 9, 13, 32],
[1, 3, 10, 15, 32],
[1, 8, 15, 33],
[1, 9, 14, 33],
[1, 18, 24, 28, 35],
[1, 18, 26, 27, 34],
[1, 19, 25, 28, 35],
[1, 20, 25, 27, 34],
[2, 8, 13, 32],
[2, 10, 14, 32],
[2, 19, 26, 27, 35],
[2, 20, 24, 27, 35],
[3, 6, 21, 27],
[3, 17, 23, 27],
[4, 5, 32, 35],
[4, 6, 12, 32, 35],
[4, 6, 22, 27],
[4, 16, 23, 27],
[5, 6, 8, 24, 29],
[5, 9, 17, 26, 30],
[5, 11, 32, 34],
[6, 7, 10, 25, 30],
[6, 8, 17, 25, 31],
[6, 10, 16, 26, 29],
[6, 11, 33, 35],
[7, 9, 16, 24, 31],
[7, 12, 32, 34],
[7, 33, 35],
[11, 13, 18, 23, 30],
[11, 15, 19, 22, 31],
[12, 14, 19, 23, 29],
[12, 15, 20, 21, 30],
[13, 20, 22, 29],
[14, 18, 21, 31],
[16, 21, 28],
[17, 22, 28]]. \\

A $6$-qubit magic assignment is \\ \\
1: IIIIIZ, 2: IIIIZI, 3: IIIZII, 4: IIIIIX, 5: IIIXIX, 6: IIIIXX, 7: IIZXII, 8: IZIZZZ, 9: IZIIII, 10: IZZIZZ, 11: IIZXXX, 12: IIIXXI, 13: ZIIZIZ, 14: ZIZIIZ, 15: ZIZZZI, 16: IIXZIX, 17: IIXIII, 18: XXYXYI, 19: XXXIIZ, 20: XXZXIZ, 21: IXXIXX, 22: IXXZXI, 23: IXIIII, 24: YZYZZZ, 25: YZXYXI, 26: YZIYZZ, 27: IXXZII, 28: IXIZXI, 29: YIYXXI, 30: YIXZZY, 31: YIIXZY, 32: ZZIIII, 33: ZZZIII, 34: ZZZIXI, 35: ZZIXII\\

To obtain MS6-35 from HA we apply the following operations. We delete vertices $\{4, 5, 7, 8, 17\}$. We then identify vertices according to the following map illustrating preimages: \\ \\
{1: \{1, 3\}, 2: \{2\}, 3: \{6\}, 4: \{9\}, 5: \{10\}, 6: \{11, 23\}, 7: \{12\}, 8: \{13\}, 9: \{14\}, 10: \{15\}, 11: \{16\}, 12: \{18\}, 13: \{19\}, 14: \{20\}, 15: \{21\}, 16: \{22\}, 17: \{24\}, 18: \{25\}, 19: \{26\}, 20: \{27\}, 21: \{28\}, 22: \{29\}, 23: \{30\}, 24: \{31\}, 25: \{32\}, 26: \{33\}, 27: \{34, 36\}, 28: \{35\}, 29: \{37\}, 30: \{38\}, 31: \{39\}, 32: \{40, 41\}, 33: \{42\}, 34: \{43\}, 35: \{44, 45\}}. \\

Notably, MS6-35 has the fewest number of observables and contexts over magic sets requiring $6$ qubits that we have found thus far.


\subsection{From HB: MS3-29}


The hypergraph HB is the following structure with $35$ observables and $35$ contexts: \\

[[1, 2, 22, 28], [1, 5, 17, 31], [1, 11, 18, 35], [1, 12, 24, 29], [2, 4, 5, 34], [2, 10, 11, 33], [2, 16, 24, 30], [3, 6, 14, 19], [3, 7, 8, 15], [3, 9, 21, 23], [3, 25, 27, 32], [4, 8, 9, 10], [4, 15, 16, 21], [4, 20, 28, 31], [5, 9, 11, 27], [5, 21, 24, 32], [6, 7, 13, 26], [6, 12, 18, 25], [6, 23, 29, 35], [7, 10, 16, 23], [7, 25, 30, 33], [8, 19, 20, 26], [8, 27, 33, 34], [9, 19, 31, 35], [10, 26, 28, 35], [11, 23, 24, 25], [12, 13, 22, 30], [12, 14, 17, 32], [13, 14, 15, 20], [13, 16, 28, 29], [14, 21, 29, 31], [15, 30, 32, 34], [17, 18, 19, 27], [17, 20, 22, 34], [18, 22, 26, 33]]. \\

In total, we have obtained $309$ magic sets from HB. In this case, the list is exhaustive.

One of the magic sets obtained from HB, the one called MS3-29 is the following structure with $29$ observables and $33$ contexts: \\

[[1, 2, 17, 22], [1, 5, 13, 25], [1, 5, 14, 29], [1, 10, 19, 23], [2, 4, 5, 28], [2, 5, 9, 27], [2, 12, 19, 24], [3, 5, 16, 18], [3, 5, 20, 26], [3, 6, 11, 15], [3, 7, 8, 11], [4, 5, 8, 9], [4, 11, 12, 16], [4, 11, 22, 25], [5, 8, 27, 28], [5, 13, 14, 15], [5, 15, 25, 29], [5, 16, 19, 26], [5, 18, 19, 20], [6, 7, 11, 21], [6, 10, 14, 20], [6, 18, 23, 29], [7, 9, 12, 18], [7, 20, 24, 27], [8, 11, 15, 21], [9, 21, 22, 29], [10, 11, 13, 26], [10, 11, 17, 24], [11, 12, 22, 23], [11, 13, 17, 28], [11, 16, 23, 25], [11, 24, 26, 28], [14, 17, 21, 27]]. \\

A $3$-qubit assignment is \\ \\
{1: IIZ, 2: IZI, 3: IZZ, 4: IIX, 5: ZII, 6: XXX, 7: XYY, 8: ZZI, 9: IZX, 10: XXI, 11: YYX, 12: XZX, 13: IXZ, 14: IXI, 15: ZIZ, 16: ZXX, 17: XIZ, 18: IYY, 19: ZZZ, 20: IXX, 21: YXY, 22: XZI, 23: YYI, 24: YZY, 25: ZXI, 26: ZYY, 27: ZIX, 28: ZZX, 29: ZXZ}. \\

To obtain MS3-29 from HB, we apply the following operations. We do not delete any vertices.
We then identify vertices according to the following map illustrating preimages: \\ \\
1: \{1\}, 2: \{2\}, 3: \{3\}, 4: \{4\}, 5: \{5, 11, 9, 27\}, 6: \{6\}, 7: \{7\}, 8: \{8\}, 9: \{10\}, 10: \{12\}, 11: \{20, 13, 14, 15\}, 12: \{16\}, 13: \{17\}, 14: \{18\}, 15: \{19\}, 16: \{21\}, 17: \{22\}, 18: \{23\}, 19: \{24\}, 20: \{25\}, 21: \{26\}, 22: \{28\}, 23: \{29\}, 24: \{30\}, 25: \{31\}, 26: \{32\}, 27: \{33\}, 28: \{34\}, 29: \{35\}. \\

We then reduce the number of occurrences of $5$ and $11$ in the hyperedges $[5,5,5,5]$ and $[11,11,11,11]$ respectively to obtain empty hyperedges. Finally, deleting the empty hyperedges, we obtain MS3-29.

Notably, the noncontextual bound of MS3-29 is $b=19$, so that its tolerance to noise is $0.424$, which is higher than the one of the pentagram, and the highest of all the magic sets that we have found.


\subsection{From HB: MS5-26}


MS5-26 has the following structure with $26$ observables and $30$ contexts: \\

[[1, 3, 11, 24], [1, 6, 18, 22], [1, 12, 26], [1, 16, 21], [2, 3, 15, 17], [2, 4, 8, 13], [2, 5, 9], [2, 19, 25], [3, 5, 10, 17], [3, 5, 19, 23], [3, 9, 23, 25], [3, 11, 14, 16], [3, 12, 16, 20], [3, 13, 24, 26], [3, 15, 18, 25], [3, 20, 21, 26], [4, 5, 7, 20], [4, 6, 12, 19], [4, 17, 22, 26], [6, 7, 16, 23], [6, 8, 11, 25], [7, 8, 9, 14], [7, 10, 21, 22], [8, 15, 22, 24], [9, 10, 15], [10, 18, 23], [11, 12, 13], [13, 14, 20], [14, 21, 24], [17, 18, 19]]. \\

A $5$-qubit magic assignment is \\ \\
{1: IIIIZ, 2: IIIIX, 3: IIIZI, 4: IIIXI, 5: IIZII, 6: IIXXI, 7: IZIXI, 8: ZIIXX, 9: IIZIX, 10: XXZZI, 11: IYXII, 12: ZYXII, 13: ZIIII, 14: ZZZII, 15: XXIZX, 16: ZXYZI, 17: XXIII, 18: YZIII, 19: ZYIII, 20: IZZII, 21: ZXYZZ, 22: YZXXZ, 23: ZYZZI, 24: IYXZZ, 25: ZYIIX, 26: ZYXIZ}. \\

To obtain MS5-26 from HB, we apply the following operations. 
We delete vertices $\{2, 4, 8, 11, 27\}$.
We then identify vertices according to the following map illustrating preimages:\\ \\
{1: \{1\}, 2: \{3\}, 3: \{33, 34, 5, 9, 10\}, 4: \{6\}, 5: \{7\}, 6: \{12\}, 7: \{13\}, 8: \{14\}, 9: \{15\}, 10: \{16\}, 11: \{17\}, 12: \{18\}, 13: \{19\}, 14: \{20\}, 15: \{21\}, 16: \{22\}, 17: \{23\}, 18: \{24\}, 19: \{25\}, 20: \{26\}, 21: \{28\}, 22: \{29\}, 23: \{30\}, 24: \{31\}, 25: \{32\}, 26: \{35\}} \\

We then reduce the multiplicity of each vertex, hyperedge pair modulo 2 and delete empty hyperedges to obtain MS5-26.

MS5-26 is notable since it has the smallest number of both observables and contexts over all $5$-qubit minimal structures found.


\subsection{From HC: MS4-21b}


The hypergraph HC is the following structure with $39$ observables and $39$ contexts: \\

[[1, 2, 8, 17], [1, 3, 9, 19], [1, 23, 24, 28], [1, 25, 26, 32], [2, 3, 16, 20], [2, 12, 34, 35], [2, 22, 23, 37], [3, 5, 26, 27], [3, 13, 34, 36], [4, 5, 10, 19], [4, 6, 11, 24], [4, 25, 27, 33], [4, 29, 30, 35], [5, 6, 21, 22], [5, 13, 38, 39], [6, 8, 28, 30], [6, 17, 37, 38], [7, 8, 14, 24], [7, 9, 15, 25], [7, 28, 29, 36], [7, 31, 32, 39], [8, 9, 23, 26], [9, 10, 32, 33], [10, 11, 27, 30], [10, 12, 18, 25], [11, 12, 16, 29], [11, 14, 35, 36], [12, 31, 33, 37], [13, 14, 20, 29], [13, 15, 21, 31], [14, 15, 28, 32], [15, 18, 37, 39], [16, 17, 23, 34], [16, 18, 33, 35], [17, 18, 22, 31], [19, 20, 26, 34], [19, 21, 27, 38], [20, 21, 36, 39], [22, 24, 30, 38]]. \\


Thus far, we have obtained $7368$ minimal configurations from HC.
One of the minimal configurations obtained from HC, called MS4-21b, is the following structure with $21$ observables and $16$ contexts: \\ 

[[1, 4, 9, 15], [1, 10, 14, 16], [2, 3, 6, 14], [2, 15, 18, 20], [3, 5, 16, 17], [3, 10, 12, 20], [3, 12, 13], [4, 5, 8, 14], [4, 6, 7, 16], [4, 18, 19, 21], [7, 9, 12, 18], [8, 9, 16, 19], [9, 11, 20, 21], [10, 11, 13, 18], [10, 13, 20], [12, 13, 14, 17]]. \\

A $4$-qubit assignment is \\ \\
1: IIIZ, 2: IIIX, 3: IIZI, 4: IIXZ, 5: IZII, 6: ZIZX, 7: IZYY, 8: ZZXZ, 9: XXXI, 10: IZIZ, 11: XXIX, 12: IYZY, 13: IYIY, 14: ZIII, 15: XXII, 16: ZZII, 17: ZIZI, 18: XIII, 19: XXIZ, 20: IXIX, 21: IXXI.\\

To obtain MS4-21b from HC, we apply the following operations. We delete the vertices $\{2, 5, 10, 20, 26, 35\}$. We then identify vertices according to the following map: \\ \\
1: \{1, 9\}, 2: \{3, 4, 12, 16, 19, 27, 34\}, 3: \{6\}, 4: \{7\}, 5: \{8\}, 6: \{11, 29\}, 7: \{13, 36\}, 8: \{14\}, 9: \{15\}, 10: \{17, 23\}, 11: \{18\}, 12: \{21, 38\}, 13: \{22\}, 14: \{24\}, 15: \{25, 33\}, 16: \{28\}, 17: \{30\}, 18: \{31\}, 19: \{32\}, 20: \{37\}, 21: \{39\}. \\ 

Reducing the multiplicity of each vertex, hyperedge pair modulo~$2$, then reducing the number of occurrences of each hyperedge modulo $2$ and finally deleting all repeated hyperedges, we obtain MS4-21b.

\begin{center}
\begin{figure*}
\includegraphics[width=0.8 \textwidth]{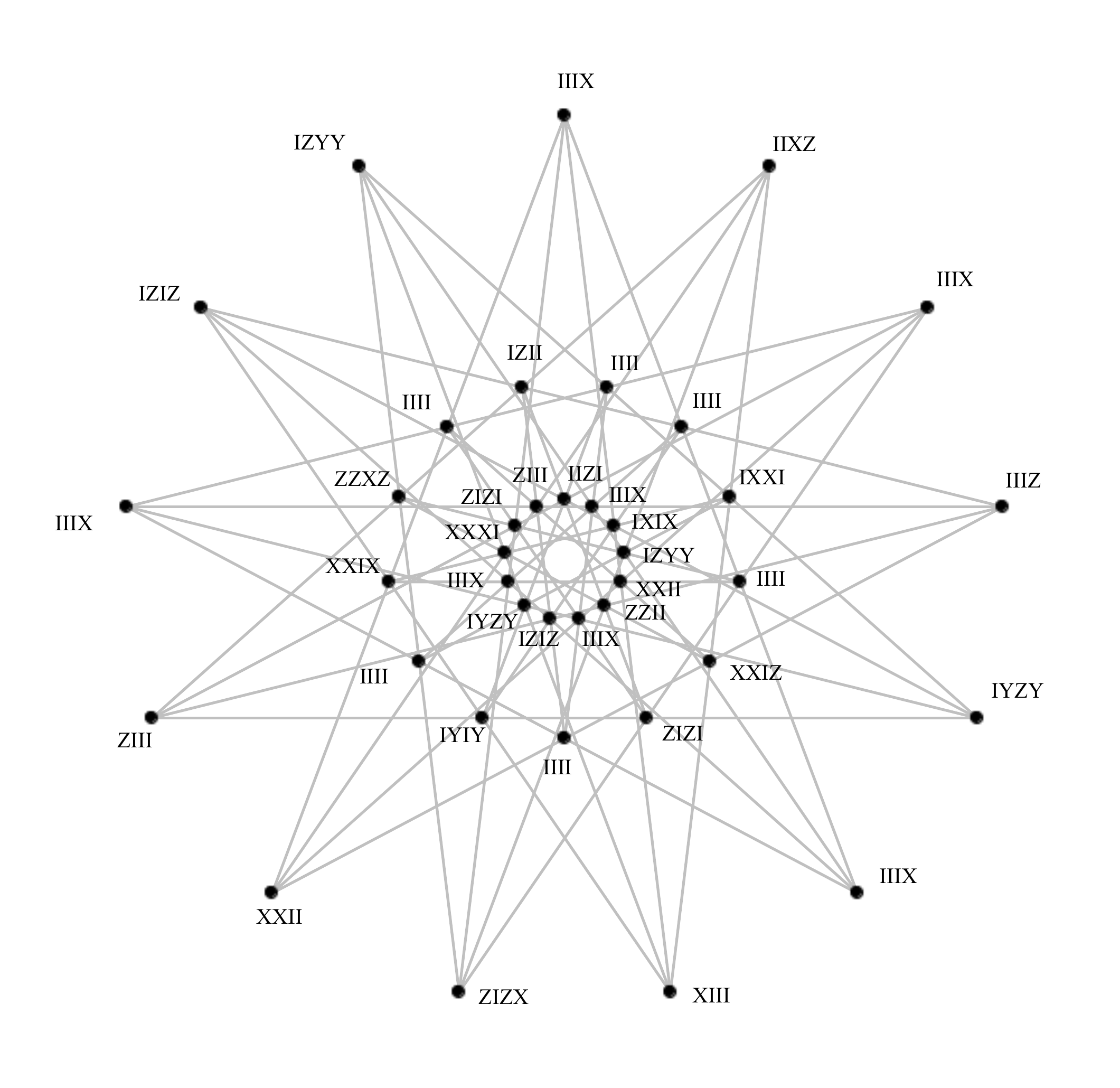}
\caption{A magic assignment of MS4-21b. The notation is explained in Fig.~1 of the main text.}
\label{fig4}
\end{figure*}
\end{center}

Fig.~\ref{fig4} illustrates the given assignment of MS4-21b via an assignment of the vertices of HC. All vertices assigned IIII are deleted, and all sets of vertices with the same assigned value are identified.

Notably, it has the fewest number of contexts over all $4$-qubit minimal structures that we have found.


\subsection{From HD: MS3-27b}


The hypergraph HD is the following structure with $45$ observables and $45$ contexts: \\

[[33, 2, 6, 19], [23, 25, 7, 40], [26, 37, 7, 10], [12, 24, 39, 40], [34, 3, 16, 20], [11, 14, 40, 30], [36, 39, 8, 20], [13, 18, 31, 43], [22, 38, 41, 10], [23, 34, 6, 8], [45, 13, 27, 41], [22, 37, 6, 21], [33, 1, 13, 28], [11, 45, 26, 29], [2, 15, 30, 32], [33, 34, 4, 18], [14, 25, 42, 43], [23, 4, 39, 19], [13, 29, 42, 10], [45, 3, 18, 30], [2, 35, 18, 21], [22, 27, 8, 42], [36, 16, 6, 32], [12, 27, 30, 43], [44, 25, 28, 10], [44, 15, 26, 40], [22, 36, 5, 9], [24, 26, 41, 9], [44, 14, 17, 32], [1, 4, 20, 31], [3, 5, 21, 32], [35, 37, 19, 9], [35, 5, 17, 31], [34, 38, 7, 21], [24, 35, 4, 7], [11, 27, 39, 9], [24, 5, 38, 20], [1, 36, 17, 19], [12, 15, 28, 41], [3, 14, 29, 31], [33, 45, 15, 16], [12, 25, 38, 8], [11, 23, 37, 42], [44, 1, 16, 29], [2, 17, 28, 43]] \\


MS3-27b is the following structure with $27$ observables and $27$ contexts: \\

[[1, 4, 13, 21], [1, 7, 10, 25], [1, 10, 14, 24], [1, 16, 21, 26], [2, 6, 15, 20], [2, 8, 11, 26], [2, 11, 13, 23], [2, 18, 20, 27], [3, 5, 14, 19], [3, 9, 12, 27], [3, 12, 15, 22], [3, 17, 19, 25], [4, 7, 18, 23], [4, 12, 19, 23], [4, 13, 17, 27], [5, 9, 16, 24], [5, 11, 20, 24], [5, 14, 18, 26], [6, 8, 17, 22], [6, 10, 21, 22], [6, 15, 16, 25], [7, 11, 17, 20], [7, 15, 22, 26], [8, 12, 16, 19], [8, 14, 24, 27], [9, 10, 18, 21], [9, 13, 23, 25]].\\

A $3$-qubit assignment is \\ \\
{1: IIZ, 2: IIX, 3: IZY, 4: ZIZ, 5: ZZY, 6: ZIX, 7: ZZZ, 8: ZZX, 9: ZIY, 10: XXZ, 11: XYI, 12: XXX, 13: IYI, 14: IXZ, 15: IXX, 16: XZZ, 17: XIY, 18: XZX, 19: ZXZ, 20: ZXX, 21: ZYI, 22: XZY, 23: XIX, 24: XIZ, 25: YYZ, 26: YXI, 27: YYX}. \\

The preimages of the vertices of MS3-27b are \\ \\
1: \{1, 10\}, 2: \{2, 12\}, 3: \{11, 3\}, 4: \{4, 13\}, 5: \{5, 14\}, 6: \{6, 15\}, 7: \{7\}, 8: \{8\}, 9: \{9\}, 10: \{16, 26\}, 11: \{17, 25\}, 12: \{18, 27\}, 13: \{19, 28\}, 14: \{20, 29\}, 15: \{21, 30\}, 16: \{22\}, 17: \{23\}, 18: \{24\}, 19: \{31, 42\}, 20: \{32, 40\}, 21: \{33, 41\}, 22: \{34, 45\}, 23: \{35, 43\}, 24: \{36, 44\}, 25: \{37\}, 26: \{38\}, 27: \{39\}. \\ 

In this case, there are no vertex, hyperedge pairs with multiplicity $\geq 1$, so we simply reduce the number of occurrences of multiple hyperedges modulo $2$ to obtain MS3-27b.

Notably, MS3-27b is the only minimal Eulerian hypergraph obtainable by reductions from HD.



\end{document}